\documentclass[12pt,round]{article}
\usepackage[latin9]{inputenc}
\usepackage[table]{xcolor}
\usepackage{colortbl}
\usepackage{array}
\usepackage{float}
\usepackage{mathrsfs}
\usepackage{multirow}
\usepackage{amsmath}
\usepackage{graphicx}
\usepackage[a4paper]{geometry}
\usepackage{setspace}
\makeatletter

\providecommand{\tabularnewline}{\\}

\makeatother

\begin{document}
\title{Solving Nash Equilibria in Nonlinear Differential Games for Common-Pool
Resources \thanks{Cai acknowledges support from the National Science Foundation grant
RISE-2108917. The authors declare that they have no other known competing
financial interests or personal relationships that could have appeared
to influence the work reported in this paper.}}
\author{Yongyang Cai\thanks{Department of Agricultural, Environmental, and Development Economics,
The Ohio State University, Columbus, USA, cai.619@osu.edu}\\
Anastasios Xepapadeas\thanks{Department of Economics, University of Bologna, Italy, and Athens
University of Economics and Business, Greece, anastasio.xepapadeas@unibo.it}\\
Aart de Zeeuw\thanks{Department of Economics, Tilburg University, the Netherlands, and
Beijer Institute for Ecological Economics, Royal Swedish Academy of
Sciences, A.J.deZeeuw@tilburguniversity.edu}}
\maketitle
\begin{abstract}
Many resources are provided by ecological systems that are vulnerable
to a sudden big loss of ecosystem services when exceeding a certain
level of pollution. This leads to non-convexities in managing ecological
systems. An ecological system is often also a common-pool resource
and therefore vulnerable to suboptimal use resulting from non-cooperative
behavior. An analysis requires methods to derive cooperative and non-cooperative
solutions in managing these types of ecological systems. Such a game
is a differential game that has two well-defined non-cooperative solutions,
the open-loop and feedback Nash equilibria. This paper provides new
numerical methods for solving open-loop and feedback Nash equilibria,
for one-dimensional and multiple-dimensional systems. The methods
are applied to the lake game, which is the classical example for these
types of problems. Especially, the two-dimensional feedback Nash equilibria
for the lake game are a novelty of this paper. Such a Nash equilibrium
can be close to the cooperative solution which has important policy
implications.



\pagebreak{}
\end{abstract}

\section{Introduction}

An important characteristic of managing common-pool resources embedded
within ecological systems\textemdash whether in the context of pollution
control or biomass harvesting\textemdash is that individual actions
of the economic agents, such as emissions or resource extraction,
generally result in accumulation of pollution stocks or depletion
of resource availability and that each agent is affected by the future
stocks emerging from all the combined individual actions (see, e.g.,
Fischer and Mirman, 1992; Dutta and Sundaram, 1993; Mäler et al.,
2003; Crépin and Lindahl, 2009). These impacts are typically damages
from the pollution stocks, increase in harvesting costs, or loss of
ecosystem services.

When each agent takes into account the impact of future stocks on
its own well-being only, the outcome is excess pollution or resource
overexploitation relative to an outcome in which the aggregate well-being
of all agents is considered. Typical examples associated with common-pool
or open access resources are climate change with excess carbon emissions
(e.g., Tahvonen, 1994; Bahn and Haurie, 2016), phosphorus pollution
of closed water bodies or closed seas (e.g., the semi-enclosed Baltic
sea (Iho et al., 2023)), or overfishing in open access fisheries (e.g.,
the Northern cod fishery in Newfoundland (Harris, 2013)).

The management of these issues involves the integration of economic
mechanisms with natural processes describing the evolution of ecosystems
which are affected by the actions of economic agents. This usually
involves two approaches. In the first, the objective is to maximize
the aggregate benefits of all economic agents subject to the constraints
emerging from the evolution of the ecosystems. This is the optimal
management approach that leads to an outcome which can be regarded
as the welfare-maximizing cooperative solution to the management problem.
In contrast, in the second approach each agent maximizes own benefits
by considering the constraints associated with the evolution of the
ecosystems and taking into account the actions of the other agents
as they seek to optimize their own benefits. This is the noncooperative
solution to the management problem.

The solution to the optimal management problem may be obtained by
standard optimal control or dynamic programming methods (see, e.g.,
Rust, 1996; Judd, 1998; Ljungqvist and Sargent, 2000; Miranda and
Fackler, 2002; Bertsekas, 2005; Bertsekas 2007; Cai, 2019) while a
noncooperative problem may be solved in the context of differential
games (Ba\c{s}ar and Olsder, 1982; Dockner et al, 2000). The most
often used solution concepts for the differential games are the weakly
time-consistent open-loop Nash equilibrium (OLNE) and the strongly
time-consistent feedback Nash equilibrium (FBNE) which possesses the
Markov perfect property, where OLNE is a solution of a time path of
agents' strategies that depend on only the initial state and time,
and FBNE is a solution of agents' strategies that depend on only the
current-period state. OLNE solutions are obtained by using optimal
control. FBNE solutions or Markov-perfect equilibria (Maskin and Tirole,
2001) are obtained in a dynamic programming context (Mäler et al,
2003; Kossioris et al, 2008; Dockner and Wagener, 2014; van der Ploeg
and Zeeuw, 2016; Gopalakrishnan et al., 2017; Jaakkola and Wagener,
2023).\footnote{For discrete time dynamic games, OLNE may be numerically solved with
an iterative method (see, e.g., Cai et al., 2023a; Cai et al., 2023b),
and FBNE may be numerically solved with a time backward iteration
method (see, e.g., Cai et al., 2019).}The standard procedure (in the case of symmetry) is to differentiate
the value function of the associated Hamilton-Jacobi-Bellman equation
with respect to the state variables which yields, by using the optimality
condition for the control, an ordinary differential equation in the
feedback equilibrium strategy as a function of the state variables.
The solution of this differential equation gives the Markov-perfect
equilibrium strategy. An important result is that even for linear-quadratic
differential games, a multiplicity of non-linear solutions exists
besides the linear solution (Tsutsui and Mino, 1990; Dockner and Long,
1993; Yanase, 2010; Nkuiya, 2015; Nkuiya and Plantinga, 2021). Dockner
and Long (1993) claim that the tragedy of the commons (in the sense
that a non-cooperative equilibrium yields lower welfare than the cooperative
outcome) is mitigated by using the non-linear strategies. However,
Wirl (2007) shows that to get this result, the elasticity of marginal
utility has to be increasing. Note that the standard procedure assumes
differentiability of the value function. Jaakkola and Wagener (2023)
derive solutions with non-smooth value functions and discontinuous
Markov-perfect equilibrium strategies and apply this to the linear-quadratic
differential game of pollution control.

An important characteristic of these problems, and one that imposes
major challenges to their solution, is the fact that the appropriate
description of ecosystem dynamics could include nonlinear feedbacks.
A prototype management problem that has been used as a vehicle for
studying cooperative and noncooperative solutions under nonlinear
ecosystem dynamics, positive feedbacks and strategic interactions
is the so-called lake problem (Carpenter and Cottingham, 1997; Scheffer,
1997).

In the lake ecosystem, economic agents benefit from emitting phosphorus\footnote{Benefits can be associated with agricultural production which generates
phosphorus emissions as runoff. The phosphorus runoff is accumulated
in the lake.} into a lake but the accumulation of the phosphorus stock generates
collective damages. The positive nonlinear feedback emerges from the
release of phosphorus from sediment at the bottom of the lake that
has accumulated as mud.\footnote{In the study of climate change, positive feedbacks could emerge from
the loss of sea ice that reduces the Earth\textquoteright s albedo,
or the thawing of permafrost that releases CO2 and methane.} The problem can be analyzed as either a one-dimensional problem in
which there is one transition equation describing the evolution of
the phosphorus stock in the lake with positive feedbacks emerging
from a fixed amount of mud, or a two-dimensional problem in which
the transition dynamics for both the phosphorus and the mud stocks
are included.

The economic management of the lake problem has mainly been analyzed
in one dimension in which the optimal management and OLNE solutions
were obtained using optimal control methods, while the FBNE problems
were solved in a dynamic programming framework (Mäler et al, 2003;
Kossioris et al, 2008; Dockner and Wagener, 2014). Optimal management
and OLNE solutions for the two-dimensional problem were acquired by
Grass et al. (2017), but to the best of our knowledge there has been
no FBNE solution for the two-dimensional lake problem. FBNE solutions
for the one-dimensional differential lake game have been obtained
in the dynamic programming context by using the standard procedure
and by solving a nonlinear differential equation in the feedback equilibrium
strategy (Kossioris et al., 2008). However, this approach only provides
a \textquotedblleft partial solution\textquotedblright{} of the problem,
because it does not provide equilibrium strategies that are defined
on the whole state space. By using a different approach, Dockner and
Wagener (2014) show that a solution exists, with discontinuous Markov-perfect
equilibrium strategies, that is defined on the whole state space.

The main contribution of this paper is the development of a novel
numerical approach, called the strategy function-value function (SFVF)
iteration, for computing FBNE solutions in differential games. For
a linear-quadratic differential game, its analytical FBNE solution
is often available when the value function is assumed to be smooth
(e.g., Dockner and Long (1993); Yanase (2010); Nkuiya (2015); Nkuiya
and Plantinga (2021)). However, solving FBNE of a general differential
game has to rely on computational methods. For stationary dynamic
games with smooth value or strategy functions, FBNEs can be computed
numerically using collocation or finite difference methods applied
to the associated Euler equations or Hamilton-Jacobi-Bellman equations
(see, e.g., Rui and Miranda, 1996; Jaakkola and van der Ploeg, 2019;
Zhu et al., 2025). However, when the strategy function is discontinuous,
the corresponding Euler or Hamilton-Jacobi-Bellman equations provide
only necessary, but not sufficient, conditions for optimality. In
such cases, the presence of multiple solutions to these equations
makes collocation and finite difference methods difficult to apply
for computing the FBNE.\footnote{Wirl (2007) delineated conditions under which such multiplicity arises
for a one dimensional pollution game, where the utility from emissions
is a power function, the transition of the pollution stock is linear,
and the external cost from the pollution stock is quadratic. The numerical
solution of Wirls\textquoteright{} (2007) problem with the use of
the SFVF method is presented in Appendix B.}

An alternative approach is to discretize the continuous-time differential
game into a discrete-time dynamic game and solve for the FBNE using
value function iteration or policy function iteration based on the
associated Bellman or Euler equations (see, e.g., Maskin and Tirole,
2001; Manzanares et al., 2015; Cai et al., 2018, 2019; Aguirregabiria
et al., 2021). This approach has been successfully applied to FBNE
computation in resource management problems (see, e.g., Balbus et
al., 2020).\footnote{Value function iteration or policy function iteration is widely used
for solving single-agent resource management problems, see, e.g.,
Knapp and Baerenklau (2006), Carlson et al. (2007), Anderson et al.
(2018), and Leach and Mason (2024).} However, if the value function is continuous but neither concave
nor convex, numerical optimization within value function iteration
may fail to identify the globally optimal solution, while the Euler
equation used in policy function iteration may encounter multiple
solutions. These issues can prevent convergence to the FBNE. Moreover,
even when all state and decision variables are discretized, the presence
of multiple equilibria can cause value function iteration to fail
to converge to the FBNE.\footnote{While value function iteration with discrete state and decision variables
can converge for a single-agent dynamic model, its convergence to
a FBNE solution cannot be guaranteed for a multiple-agent dynamic
game, because one agent's decision depends on other agents' strategies
too. In addition, the discretization implies that we cannot use an
Euler equation for policy function iteration.} In addition, the curse-of-dimensionality inherent in discretization
severely limits the applicability of these methods in settings with
multiple state or decision variables.

The SFVF method accommodates non-smooth value functions and is based
on joint iteration of the Markov-perfect strategy function and the
value function, using the Hamilton-Jacobi-Bellman equation together
with the associated first-order optimality conditions. For one-dimensional
or multi-dimensional nonlinear differential games in which the value
function associated with the FBNE is smooth and there are no multiple
equilibria, existing numerical approaches\textemdash including collocation
methods, finite difference methods, value function iteration, and
policy function iteration\textemdash can be effective in computing
the FBNE. However, the value function could be non-concave and non-convex,
and it might not be smooth at some points in the state space. Under
these conditions, the associated Hamilton-Jacobi-Bellman equation
may admit multiple solutions, and existing methods often fail to compute
the FBNE. In such settings, the proposed SFVF approach may provide
a feasible and efficient method for computing the FBNE.

Applying the SFVF method to the lake problem, we reproduce the FBNE
solution on the entire state space for the one-dimensional model and,
to our knowledge, obtain for the first time solutions for the two-dimensional
model. The same approach is used to reproduce the optimal management
solutions for both the one-dimensional and two-dimensional lake problems.
Moreover, we introduce a new and efficient numerical method for solving
the boundary value problem emerging in the computation of the OLNE.

In this paper, the lake problem serves as a vehicle for demonstrating
the ability of the SFVF method to solve nonlinear dynamic optimization
problems. More generally, the approach is applicable to nonlinear
resource management problems with strategic interactions and multiple
state and control variables, including predator-prey open-access fishery
management problems, grazing in common-pool grasslands, and common-property
aquifer management.

The rest of the paper is organized as follows. Section 2 introduces
the SFVF method and the numerical method for solving boundary value
problems. Section 3 presents the model of the lake ecosystem in the
one-dimensional and two-dimensional representations. Sections 4 and
5 provide the SFVF algorithms for the solution of the one-dimensional
and two-dimensional lake problems respectively, along with the corresponding
solutions for optimal management, OLNE and FBNE. Section 6 provides
a summary and discussion of the results and Section 7 concludes.

\section{New Methods for Obtaining Nash Equilibrium Solutions}

Without loss of generality, here we assume each agent $a$ has only
one decision variable $x_{a}$ for convenience. In a general continuous
time dynamic game with $n$ homogeneous agents, we assume one agent
$a$ has a utility function $u(x_{a}(t),\mathbf{S}(t))$ at time $t$,
and $\mathbf{S}$ is the state variable vector for all agents. An
open-loop Nash equilibrium (OLNE) assumes that each agent's time path
of their strategies depends on only the initial state vector $\mathbf{S}_{0}$
and time $t$. A feedback Nash equilibrium (FBNE) assumes that each
agent's strategy depends on only the current-period state variable
vector $\mathbf{S}$(t) at time $t$. When $n=1$, the lake problem
as formulated by Kossioris et al. (2008) or Grass et al. (2017) is
degenerated into an optimal management problem, which may also be
solved with the following methods for OLNE or FBNE. In this paper,
we assume that symmetric equilibria exist in our models, and our numerical
methods are designed for solving the symmetric equilibria.

\subsection{Method for OLNE\label{subsec:Method-for-OLNE}}

An OLNE for agent $a$ is solving
\[
\max_{x_{a}(\cdot)}\int^{\infty}_{0}e^{-\rho t}u(x_{a}(t),\mathbf{S}(t))dt
\]
with the following transition law 
\[
\dot{\mathbf{S}}(t)=\mathbf{F}(x(t),\mathbf{S}(t))
\]
with $\mathbf{S}(0)=\mathbf{S}_{0}$, where $x(t)$ is a function
of $x_{a}(t)$ and all other agents' decision vectors: $x_{a'}(t)$
for every $a'\neq a$, denoted $x=X(x_{a},x_{a'})$. In our lake example,
$x_{a}$ is the phosphorus loading $L_{a}$ by agent $a$, $x$ is
the total loading (i.e., $x(t)=L_{a}(t)+\sum_{a'\neq a}L_{a'}(t)$),
and $\mathbf{S}=(P,M)$ with $P$ the phosphorus density in the water
of the lake and $M$ the phosphorus density in the sediment of the
lake. Its associated Hamiltonian function is 
\[
H(x_{a},\mathbf{S},\mathbf{y},t)=e^{-\rho t}u(x_{a},\mathbf{S})+\mathbf{y}'\mathbf{F}(x,\mathbf{S}),
\]
where $\mathbf{y}$ is the vector of co-states, and the associated
Hamiltonian system is 
\begin{eqnarray*}
\dot{\mathbf{S}} & = & \mathbf{F}(x,\mathbf{S})\\
\frac{\partial H}{\partial x_{a}} & = & 0\\
\dot{\mathbf{y}} & = & -\frac{\partial H}{\partial\mathbf{S}}
\end{eqnarray*}
while the transversality condition depends on the model setting. For
our lake example, the transversality condition is
\[
\lim_{t\rightarrow\infty}\mathbf{y}(t)=0
\]
which can be satisfied by assuming that the lake system converges
to a finite steady state. With the first-order optimality condition
on $x_{a}$ and homogeneity of decisions, the decision variable $x_{a}$
may be substituted by a function of $(\mathbf{S},\mathbf{y})$ (or
some co-states may be substituted by a function of state variables,
decision variables, and other co-states), so the Hamiltonian system
can be transformed to a system of ordinary differential equations.
In this paper, we assume that the Hamiltonian system is sufficient
for computing OLNE.

A common numerical method views the system of ordinary differential
equations as a boundary value problem with a pre-specified initial
state and the transversality condition, then apply a solver, such
as bvp4c in Matlab, to solve it (see, e.g., Gopalakrishnan et al.,
2017). However, the method might fail in solving problems with multiple
equilibria. Alternatively, the transversality condition can be replaced
by the existence of steady states in the long run, which is often
sufficient to satisfy the transversality condition, then use the steady
state as the terminal state in the boundary value problem. After solving
all steady states, this alternative method can solve problems with
multiple equilibria (see, e.g., Grass et al., 2017). That is, with
the system of ordinary differential equations, we can solve all steady
states simply by finding the solutions of the following system: 
\begin{eqnarray}
\mathbf{F}(x,\mathbf{S}) & = & 0\label{eq:ss1}\\
\frac{\partial H}{\partial x_{a}} & = & 0\label{eq:ss2}\\
\frac{\partial H}{\partial\mathbf{S}} & = & 0\label{eq:ss3}\\
x & = & X(x_{a},x_{a})\label{eq:ss4}
\end{eqnarray}
Here the second argument of (\ref{eq:ss4}) is replaced by $x_{a}$
due to the symmetric property of homogeneous agents in the Nash equilibrium,
therefore the unknown variables of the above system (\ref{eq:ss1})-(\ref{eq:ss4})
are just $x,$ $x_{a}$, $\mathbf{S}$, and $\mathbf{y}$. Therefore,
we are solving a boundary value problem with a pre-specified initial
state and a terminal state being a steady state. However, an existing
solver like bvp4c is often unstable, particularly because our system
of ordinary differential equations is an infinite horizon problem
which requires a long time horizon to converge to a steady state.

This paper introduces a simple and efficient numerical method to solve
the boundary value problem. We let $\tau=1-\exp(-\lambda t)$ to normalize
the infinite time horizon to $[0,1)$, where $\lambda$ is a small
positive number. The intuition is that when $t$ is larger, the dynamic
system is closer to its steady state, but it has a slower velocity
(as $\dot{\mathbf{S}}$ is closer to zero), then this transformation
speeds up the velocity at a large time $t$ (i.e., $\tau$ is close
to 1). Since $d\tau/dt=\lambda(1-\tau)$, the boundary value problem
can be transformed to another boundary value problem depending on
$\tau$ instead of $t$. To the best of our knowledge, this is the
first time to apply the transformation to solving the OLNE of a differential
game. Now we apply the bvp4c solver in Matlab to solve the new boundary
value problem for one starting state vector $\mathbf{S}_{0}$, where
the terminal condition is that the associated variables at a terminal
time $\mathscr{T}$ are equal to their values at a steady state vector,
where $\mathscr{T}<1$ is chosen to be close to 1. The initial guess
for the solution at one starting point can use the steady state values
or the solution of the previous starting point. When there are multiple
steady states, the Hamiltonian system might have multiple solutions,
and we choose the one with the highest welfare. Our lake examples
show that this method is successful in solving the OLNE.

\subsection{Method for FBNE\label{subsec:Method-for-FBNE}}

Let $G(\mathbf{S})$ be the time stationary strategy function of one
agent under the FBNE with $n$ homogeneous agents. For every agent
$a$, we are solving
\begin{equation}
V(\mathbf{S}_{0})=\max_{x_{a}(\cdot)}\int^{\infty}_{0}e^{-\rho t}u(x_{a}(t),\mathbf{S}(t))dt\label{eq:FBNE-model}
\end{equation}
 subject to the following transition law 
\[
\dot{\mathbf{S}}(t)=\mathbf{F}(x(t),\mathbf{S}(t))
\]
with $\mathbf{S}(0)=\mathbf{S}_{0}$, where $x(t)$ is a function
of $x_{a}(t)$ and the unknown function $\mathbf{G}(\mathbf{S}(t))$,
denoted $x=X(x_{a},G(\mathbf{S}))$, and $V$ is the value function.
In our lake example, $x_{a}$ is the loading $L_{a}$, $x$ is the
total loading (i.e., $L(t)=L_{a}(t)+(n-1)G(\mathbf{S}(t))$), and
$\mathbf{S}=(P,M)$. Its associated Hamilton-Jacobi-Bellman equation
is

\begin{equation}
\rho V\left(\mathbf{S}\right)=\max_{x_{a}}\left\{ u(x_{a},\mathbf{S})+\left(\nabla V\right)\cdot\mathbf{F}(x,\mathbf{S})\right\} \label{eq:HJB_max}
\end{equation}
where $\nabla V$ is the gradient vector of the value function $V$,
and $\left(\nabla V\right)\cdot\mathbf{F}(x,\mathbf{S})$ is the inner
product of $\nabla V$ and $\mathbf{F}(x,\mathbf{S})$. With the first-order
condition of the maximization problem, we find $V$ by solving the
following system of equations 
\begin{eqnarray}
\frac{\partial u}{\partial x_{a}}+\left(\nabla V\right)\cdot\left(\frac{\partial\mathbf{F}}{\partial x}\frac{\partial X}{\partial x_{a}}\right) & = & 0\label{eq:FOC_FBNE}\\
u(x_{a},\mathbf{S})+\left(\nabla V\right)\cdot\mathbf{F}(x,\mathbf{S}) & = & \rho V\left(\mathbf{S}\right)\label{eq:HJB_FBNE}\\
x & = & X(x_{a},x_{a})\label{eq:total_x_FBNE}
\end{eqnarray}
Here the second argument of the function $X$ in (\ref{eq:total_x_FBNE})
becomes $x_{a}$ as $x_{a}=G(\mathbf{S})$, due to the symmetric property
of homogeneous agents in the Nash equilibrium, therefore the above
system (\ref{eq:FOC_FBNE})-(\ref{eq:total_x_FBNE}) is transformed
to a differential equation in $V$ only.

However, the equations (\ref{eq:FOC_FBNE})-(\ref{eq:total_x_FBNE})
are necessary but not sufficient conditions for obtaining the feedback
Nash equilibrium. When the conditions are not sufficient, we need
to consider the original definition of $V$ in (\ref{eq:FBNE-model}).
When the value function is not smooth on some points and the equations
(\ref{eq:FOC_FBNE})-(\ref{eq:total_x_FBNE}) have multiple solutions,
as in our lake examples, it is challenging to solve the equations
(\ref{eq:FOC_FBNE})-(\ref{eq:total_x_FBNE}) with standard computational
methods like finite difference methods and projection methods (Judd,
1998). If we apply a standard value function iteration to solve the
FBNE, we need to discretize time to transform (\ref{eq:FBNE-model})
to a Bellman equation, then iterate the value function approximation
until convergence, but such iteration often fails when the value function
is non-smooth, non-convex and non-concave (as in our lake examples).
Moreover, if the time discretized solution needs to approximate the
continuous time solution, it requires a small time step size, implying
a high accuracy requirement in optimization and a slow convergence
rate if it converges.

Here we introduce a novel strategy function-value function (SFVF)
iteration method to solve the FBNE model (\ref{eq:FBNE-model}). It
is rooted in a combination of the principles of value function iteration
and policy function iteration, by updating the value function and
the policy function based on the definition of the value function
and the Hamilton-Jacobi-Bellman equation respectively. The algorithm
to solve the FBNE model (\ref{eq:FBNE-model}) is as follows:
\begin{description}
\item [{Algorithm}] 1. Strategy Function-Value Function (SFVF) Iteration
for FBNE
\item [{Step}] 1. Choose a set of points $\{\mathbf{S}_{i}:\ i=1,...,N\}$
on a pre-specified state space.\footnote{The number of points is chosen such that more points have little impact
on the solution.} Set an initial guess of the strategy function $G^{0}(\mathbf{S})$
and its associated value function $V^{0}(\mathbf{S})$. Iterate through
steps 2-4 for $j=0,1,2,...$, until convergence.
\item [{Step}] 2. Update the strategy function. Solve the following system
of equations 
\begin{eqnarray}
\frac{\partial u}{\partial x_{a}}+\mathbf{y}\cdot\left(\frac{\partial\mathbf{F}(x,\mathbf{S}_{i})}{\partial x}\frac{\partial\mathbf{X}}{\partial x_{a}}\right) & = & 0\label{eq:FOC_FBNE_Alg}\\
u(x_{a},\mathbf{S}_{i})+\mathbf{y}\cdot\mathbf{F}(x,\mathbf{S}_{i}) & = & \rho V^{j}\left(\mathbf{S}_{i}\right)\label{eq:HJB_FBNE_Alg}\\
x & = & X(x_{a},x_{a})\label{eq:total_x_FBNE_Alg}
\end{eqnarray}
to find $x_{a}$ and $\mathbf{y}$ for each $i=1,...,N$, where $\mathbf{y}$
corresponds to $\nabla V^{j}(\mathbf{S}_{i})$. When there are multiple
state variables, the number of unknowns in the equations (\ref{eq:FOC_FBNE_Alg})-(\ref{eq:total_x_FBNE_Alg})
is larger than the number of equations, so we choose one appropriate
element of $\mathbf{y}$ (if one element of $\mathbf{y}$ can be solved
directly from the equation (\ref{eq:FOC_FBNE_Alg}), then the element
is chosen) and substitute the other elements of $\mathbf{y}$ by numerical
approximation of their corresponding elements of $\nabla V^{j}(\mathbf{S}_{i})$.\footnote{See Algorithm 1.2 for an example.}
When there are multiple solutions of $x_{a}$ for one $i$, choose
the one that makes the left hand side of the equation (\ref{eq:FOC_FBNE})
be closest to zero, where $\nabla V^{j}(\mathbf{S}_{i})$ can be estimated
from a finite difference method. The solution of $x_{a}$ for $i$
is denoted $x_{a,i}$. Use an appropriate approximation method to
construct a strategy function $G^{j+1}(\mathbf{S})$ such that $G^{j+1}(\mathbf{S}_{i})=x_{a,i}$
for all $i$.
\item [{Step}] 3. Update the value function. Use $G^{j+1}(\mathbf{S})$
to generate a trajectory $(\mathbf{S}(t),x_{a}(t))$ starting at $\mathbf{S}(0)=\mathbf{S}_{i}$
by letting 
\[
x_{a}(t)=G^{j+1}(\mathbf{S}(t))
\]
and 
\[
\mathbf{S}(t+h)=\mathbf{S}(t)+\mathbf{F}(X(x_{a}(t),G^{j+1}(\mathbf{S}(t))),\mathbf{S}(t))h
\]
where $h$ is a small time step size, and use a numerical quadrature
method to estimate the welfare at $\mathbf{S}_{i}$: 
\[
v_{i}=\int^{\infty}_{0}e^{-\rho t}u(x_{a}(t),\mathbf{S}(t))dt,
\]
and use an appropriate approximation method to construct a value function
$V^{j+1}(\mathbf{S})$ such that 
\[
V^{j+1}(\mathbf{S}_{i})=\omega V^{j}(\mathbf{S}_{i})+(1-\omega)v_{i}
\]
with a parameter $\omega\in(0,1)$, for all $i$.
\item [{Step}] 4. Check if $V^{j+1}\approx V^{j}$ and $G^{j+1}\approx G^{j}$.
If so, stop the iteration, otherwise go to Step 2 by increasing $j$
with 1.
\end{description}
In Step 1 of Algorithm 1, after we have an initial guess of the strategy
function $G^{0}(\mathbf{S})$, we use the numerical quadrature approach
in Step 3 to estimate its associated welfare 
\begin{equation}
v_{i}=\int^{\infty}_{0}e^{-\rho t}u(x_{a}(t),\mathbf{S}(t))dt\label{eq:num_welfare}
\end{equation}
starting at $\mathbf{S}(0)=\mathbf{S}_{i}$, then apply an appropriate
approximation method to construct the initial guess of the value function,
$V^{0}(\mathbf{S})$, such that $V^{0}(\mathbf{S}_{i})=v_{i}$ for
all $i$. We may also adjust $V^{0}(\mathbf{S})$, e.g., by adding
one constant, as the estimation of the welfare $v_{i}$ has numerical
errors and the algorithm does not require a complete match between
$G^{0}(\mathbf{S})$ and $V^{0}(\mathbf{S})$. Similar to most of
numerical methods, the convergence of Algorithm 1 depends on the initial
guess: $G^{0}(\mathbf{S})$ and $V^{0}(\mathbf{S})$. A good way of
choosing an initial guess here is to assume that the states $\mathbf{S}$
are steady states so that the value function is given by (\ref{eq:FBNE-model})
or (\ref{eq:num_welfare}) as the net present value for the corresponding
optimal decision. For example, in our lake example, we choose an initial
guess such that $V^{0}(\mathbf{S})$ at each steady state $\mathbf{S}_{ss}$
is close to its associated true maximal welfare, $u(x_{a,ss},\mathbf{S}_{ss})/\rho$,
where $x_{a,ss}$ is the optimal decision at the steady state $\mathbf{S}_{ss}$,
and extend this to get an initial guess for the other states $\mathbf{S}$.
Note that in our lake example the value function has the property
that it is monotonically decreasing over the state variables, so we
choose $V^{0}(\mathbf{S})$ to have the same property.

In Step 2 of Algorithm 1, the equations (\ref{eq:FOC_FBNE_Alg})-(\ref{eq:HJB_FBNE_Alg})
are consistent with the equations (\ref{eq:FOC_FBNE})-(\ref{eq:HJB_FBNE}).
The equations (\ref{eq:FOC_FBNE_Alg})-(\ref{eq:total_x_FBNE_Alg})
can be solved with the fsolve package in Matlab. Using different initial
guesses of $x_{a}$ and $\mathbf{y}$, we may have different solutions
of (\ref{eq:FOC_FBNE_Alg})-(\ref{eq:total_x_FBNE_Alg}) at one state
$\mathbf{S}_{i}$, denoted $x^{k}_{a,i}$ for $k=1,...,K$, where
$K$ is the number of different solutions. We choose the one that
makes the left hand side of the equation (\ref{eq:FOC_FBNE}) be closest
to zero, where $\nabla V^{j}(\mathbf{S}_{i})$ can be estimated from
a finite difference method. Note that this process in practice requires
using a sufficiently large number of initial guesses covering the
range of admissible solutions. However, this choice is good only when
the value or strategy function is close to a FBNE. Therefore, to increase
the stability and efficiency of the algorithm, when the difference
between two consecutive value or strategy functions is not small,
we just choose the one which is closest to the corresponding $G^{j}(\mathbf{S}_{i})$
in the last iteration. In Step 3 of Algorithm 1, we can use a numerical
quadrature method with $h$ as the step size to estimate the integral,
by truncating the integral's range $[0,\infty)$ to $[0,T]$ with
a large $T$.\footnote{$T$ and $h$ are chosen such that a larger $T$ or a smaller $h$
has little impact on the results.}

In the end, it is necessary to measure whether the converged solution
is a FBNE. Since the converged value and strategy functions satisfy
equations (\ref{eq:FOC_FBNE_Alg})-(\ref{eq:total_x_FBNE_Alg}), it
remains only to verify the equivalence between $\mathbf{y}$ and $\nabla V^{j}(\mathbf{S}_{i})$,
which in turn ensures that equations (\ref{eq:FOC_FBNE})-(\ref{eq:total_x_FBNE})
are satisfied. Moreover, we should verify whether the solution $x_{a,i}$
is the global maximizer in the Hamilton-Jacobi-Bellman equation (\ref{eq:HJB_max})
at the state $\mathbf{S}_{i}$ after we substitute $V$ by the converged
value function approximation, by using a global optimization solver
(e.g., GlobalSearch in Matlab). However, in our lake problems, (\ref{eq:HJB_max})
has one unique maximizer, that is, once $\nabla V$ and the other
players' strategy function $G$ are given, (\ref{eq:FOC_FBNE}) has
one unique solution for $x_{a}$, so this step can be omitted. Since
the global maximizer in the Hamilton-Jacobi-Bellman equation (\ref{eq:HJB_max})
is based on the value function $V$ and the strategy function $G$,
we cannot derive that it is also the global maximizer of the original
model (\ref{eq:FBNE-model}). That is, Algorithm 1 may provide a FBNE
satisfying (\ref{eq:FOC_FBNE})-(\ref{eq:total_x_FBNE}), but it might
not be the one with the highest welfare, due to the potential existence
of multiple FBNEs, as shown in the literature. Similar to the well-known
difficulty of identifying all local maximizers or the global maximizer
in a general nonlinear maximization problem, it is also challenging
to find all FBNEs of differential games and to determine the FBNE
that yields the highest welfare. Addressing this challenge remains
a topic for future research.

\section{The lake model}

Limnologists study lakes and observe that the state of a lake can
suddenly change from a healthy state to an unhealthy state, with a
big loss of ecosystem services such as fish, clean water, and several
amenities (Carpenter and Cottingham, 1997, Scheffer, 1997). The cause
is the accumulation of phosphorus in the lake which results from the
release of phosphorus on the lake from surrounding agriculture. At
some point, a small increase in the phosphorus loading shifts the
lake to a state with a much lower level of ecosystem services (Scheffer
et al, 2001). Such a point is called a tipping point. The basic model
for the lake that explains these observations consists of three differential
equations: one for the accumulation of phosphorus in the water of
the lake, one for the accumulation of phosphorus in the sediment of
the lake, and one for the loading of phosphorus on the lake (Carpenter,
2005). Considering optimal management of the lake or the lake game
turns the phosphorus loadings into control variables, so that the
underlying lake model becomes a system of two differential equations
given by

\begin{equation}
\begin{cases}
\dot{P}(t)=L(t)+f(P(t),M(t)),\\
\dot{M}(t)=g(P(t),M(t)),
\end{cases}\label{eq:trans-law-2D}
\end{equation}
where $(P(0),M(0))=(P_{0},M_{0})$ is an initial state, and
\begin{eqnarray*}
f\left(P,M\right) & := & -(s+\varsigma)P+rM\frac{P^{\alpha}}{P^{\alpha}+q^{\alpha}},\\
g\left(P,M\right) & := & sP-\eta M-rM\frac{P^{\alpha}}{P^{\alpha}+q^{\alpha}},
\end{eqnarray*}
where $L$ denotes the total phosphorus loadings, $P$ the phosphorus
density in the water of the lake, and $M$ the phosphorus density
in the sediment of the lake (with $M$ for mud). The parameter $s$
denotes the sedimentation rate, $\varsigma$ the outflow rate, $\eta$
the permanent burial rate, and $r$ the maximum recycling rate. The
non-linear term is called a Holling type-III functional response term,
and this term yields the concave-convex shape of the lake equilibria
which makes tipping points possible. The model is estimated and tested
using observations on Lake Mendota, Wisconsin, USA: $s=0.7$, $\varsigma=0.15$,
$\eta=0.001$, $r=0.019$, $q=2.4$, $\alpha=8$. The small values
of $r$ and $\eta$ imply that the dynamics of the second equation
in (\ref{eq:trans-law-2D}) is much slower than the dynamics of the
first equation (Janssen and Carpenter, 1999). The literature therefore
first assumed that $M$ is constant, starting with Brock and Starrett
(2003), Mäler et al (2003) and Wagener (2003), and later moved to
an analysis with fast and slow dynamics in $P$ and $M$ (Grass et
al, 2017).

The typical common-property issue arises if $n$ economic agents have
private benefits from using the lake as a phosphorus sink but each
suffers from the damage to the lake by aggregate phosphorus loadings.
Suppose that the objective functions are given by
\begin{equation}
\max_{L_{a}(\cdot)}\int^{\infty}_{0}e^{-\rho t}\left[\ln(L_{a}(t))-cP^{2}(t)\right]dt,\ \ a=1,...,n,\label{eq:obj-2D}
\end{equation}
with the total loadings $L=\sum^{n}_{a=1}L_{a}$. The parameter $c$
denotes the relative weight of the damage compared to the benefits,
and the logarithm is convenient because the cooperative solution is
then independent of the number of agents $n$, and it is equal to
the optimal management solution. Assuming $M$ is constant, (\ref{eq:trans-law-2D})
reduces to a differential equation in $P$. In the numerical calculations
we change the power to $\alpha=2$, to simplify without affecting
the qualitative structure of the results.\footnote{Rewriting $x=P/q$, $u=L/(rM)$, $b=(s+\varsigma)q/(rM)$, and changing
the time scale to $rMt/q$ yields 
\begin{equation}
\dot{x}(t)=u(t)-bx(t)+\frac{x^{2}(t)}{x^{2}(t)+1},\ x(0)=x_{0}.\label{eq:simple_law}
\end{equation}

which is the basic equation used to describe lake dynamics in the
traditional analysis of the lake problem.}

For the one-dimensional lake game in which the lake dynamics are described
by equation (\ref{eq:simple_law}), Mäler et al (2003) compared the
cooperative solution with the open-loop Nash equilibrium where the
loading strategies $L_{a}$ depend on time $t$ only. For certain
values of the parameters the cooperative solution moves the lake to
a steady state with a high level of ecosystem services, but the open-loop
Nash equilibrium has two steady states, one with a high and one with
a low level of ecosystem services. These regions are called the oligotrophic
and the eutrophic region of the lake. The open-loop Nash equilibrium
is not unique but because of symmetry, the agents can coordinate on
the best one. It depends on the initial condition whether the lake
ends up in an oligotrophic steady state or in a eutrophic steady state.
The same occurs for the cooperative solution, or optimal management,
for a lower cost parameter $c$. The initial condition on the state
where the agents are indifferent is called a Skiba point (Skiba, 1978).

Kossioris et al (2008) and Dockner and Wagener (2014) derived, for
the one-dimensional lake game, the feedback Nash equilibrium where
the loading strategies $L_{a}$ depend on the state $P$. The literature
shows that even for a differential game with a linear state equation
and quadratic objective functionals, a multiplicity of (non-linear)
feedback Nash equilibria exist (Tsutsui and Mino, 1990; Dockner and
Long, 1993). The same occurs for the lake game. Kossioris et al (2008)
follow this literature and solve the differential equation in the
feedback strategy emerging from differentiating the Hamilton-Jacobi-Bellman
equation with respect to the state variable $P$ and using the corresponding
optimality condition. Dockner and Wagener (2014) transform the same
differential equation into a two-dimensional system. This allows them
to solve for the feedback strategy that is defined on the entire state
space, and to identify a discontinuity in the strategy function. For
the same parameter values as in the open-loop case above, the feedback
Nash equilibrium has only one steady state in the oligotrophic regime
and moves the lake out of the eutrophic regime. The interactions between
the agents through observations on the phosphorus stock allow for
a Nash equilibrium moving the lake to the oligotrophic regime. The
steady state of the feedback Nash equilibrium is close to the optimal
management steady state and converges to the optimal management steady
state if the discount rate $\rho$ goes to zero.

Grass et al (2017) analyze the two-dimensional lake game (\ref{eq:obj-2D})
subject to the dynamical system (\ref{eq:trans-law-2D}), with fast
and slow dynamics. They choose parameter values, with the fixed stock
of phosphorus in the sediment of the lake $M=179$, such that the
results can be compared to the results for the one-dimensional reduced
form (\ref{eq:simple_law}). For initial condition $M_{0}=179$, the
optimal management path for the two-dimensional lake game converges
to the steady state $(P^{*},M^{*})=(0.774,194.2)$ which is in the
oligotrophic regime of the lake. The adjustment in $P$ is fast, followed
by slow adjustments in $M$ and thus in $P$, along the isocline $\dot{P}=0$.
However, for the higher initial condition $M_{0}=240$, another type
of Skiba or indifference point occurs. This Skiba point does not separate
different domains of attraction with different steady states but separates
different optimal paths towards the same steady state. These paths
have the same optimal value. One path moves fast to the oligotrophic
regime of the lake and then slowly adjusts to the steady state. The
other path moves to the eutrophic regime first, then slowly adjusts
with $M$, and finally moves fast to the long-run steady state in
the oligotrophic regime. Optimal management becomes indifferent between
moving to the oligotrophic regime immediately or staying in the eutrophic
regime for a while and moving there later. This is called a weak Skiba
point. By gradually lowering $M_{0}$, a curve of weak Skiba points
in the $(P,M)$-plane appears but at some point, the optimal management
path always moves directly to the oligotrophic regime.

For a lower cost parameter $c$, the traditional Skiba or indifference
points return. There are two steady states, one in the oligotrophic
regime of the lake and one in the eutrophic regime. The result is
basically the same as the result for the reduced form of the lake.
A Skiba manifold in the $(P,M)$-plane appears that separates the
domains of attraction for the two steady states. For the initial conditions
$M_{0}=179$ and $P_{0}$ either to the left or the right of the Skiba
manifold, the result is approximately the same as the result for the
reduced form of the lake. The adjustment in $P$ to either the oligotrophic
or eutrophic regime of the lake is fast, followed by slow adjustments
in $M$ and thus in $P$, along the isocline $\dot{P}=0$. For higher
initial conditions $M_{0}$, the adjustment process to one of the
steady states takes longer. Lowering the parameter $c$ further yields
a bifurcation back to one steady state, but this steady state is in
the eutrophic regime of the lake. A mirror picture results as compared
to the case for a high $c$, with again a weak Skiba manifold for
high initial conditions $M_{0}$.

\section{One Dimensional Lake Problems}

In the one-dimensional lake problem, we assume that the phosphorus
density in the sediment of the lake, $M$, is constant, i.e., it exhibits
no dynamics. Under this assumption, we compute three types of solutions:
the cooperative (or optimal management) solution, the OLNE, and the
FBNE. The cooperative solution is defined as follows:

\begin{equation}
\max_{L_{\alpha}(\cdot),a=1,...n}\int^{\infty}_{0}e^{-\rho t}\left[\sum^{n}_{a=1}\ln(L_{a}(t))-ncP(t)^{2}\right]dt\label{eq:CoopObj}
\end{equation}
subject to 
\begin{equation}
\dot{P}(t)=L(t)+f(P(t)),\ P(0)=P_{0}\label{eq:tran-law-1D-coop}
\end{equation}
where 
\begin{eqnarray*}
L(t) & = & \sum^{n}_{a=1}L_{a}(t)\\
f\left(P\right) & := & -(s+\varsigma)P+rM\frac{P^{\alpha}}{P^{\alpha}+q^{\alpha}}
\end{eqnarray*}
with the constant $M$. The logarithmic form and the assumption of
agent symmetry imply that the cooperative solution is independent
of the number of agents (Grass et al. 2017).\footnote{Under symmetry the sum of utilities in (\ref{eq:CoopObj}) can be
written as $n\left[\ln(L/n)-cP(t)^{2}\right]=n\left[\ln(L)-\ln n-cP(t)^{2}\right]$.} In this study, we solve the cooperative problem using the SFVF method
and setting $n=1$. This is equivalent to solving an optimal management
problem in the total loadings $L(t).$

Since the SFVF approach is more effectively demonstrated in the context
of the FBNE, we present the cooperative solution after introducing
the FBNE results. Accordingly, this section begins with the application
of the boundary value problem method to derive the OLNE, followed
by the FBNE and the cooperative solution obtained via the SFVF method.

\subsection{Open-loop Nash Equilibrium\label{subsec:Open-loop-Nash-Equilibrium-1D}}

The open-loop Nash equilibrium with $n$ agents (where each agent
$a$ has the homogeneous utility function $\ln(L_{a}(t))-cP(t)^{2}$
at time $t$) is obtained by solving
\[
\max_{L_{a}(\cdot)}\int^{\infty}_{0}e^{-\rho t}\left[\ln(L_{a}(t))-cP(t)^{2}\right]dt
\]
subject to 
\begin{equation}
\dot{P}(t)=L(t)+f(P(t)),\ P(0)=P_{0}\label{eq:trans-law-P-1D}
\end{equation}
where 
\begin{eqnarray*}
L(t) & = & L_{a}(t)+\sum_{a'\neq a}L_{a'}(t).
\end{eqnarray*}
From its associated Hamiltonian system, the problem is equivalent
to solving the following system of ordinary differential equations:
\begin{eqnarray}
\dot{P} & = & L+f(P)\label{eq:OLNE-ODE-1D-P}\\
\dot{L} & = & \left[f'(P)-\rho\right]L+\frac{2cP}{n}L^{2}\label{eq:OLNE-ODE-1D_L}
\end{eqnarray}
This is a boundary value problem with a pre-specified initial state
and a terminal state being a steady state. It is simple to find all
steady states in OLNE by solving 
\begin{eqnarray}
0 & = & L+f(P)\label{eq:ss1-oneD}\\
0 & = & \left[f'(P)-\rho\right]+\frac{2cP}{n}L\label{eq:ss2-oneD}
\end{eqnarray}

We apply the method in Section \ref{subsec:Method-for-OLNE} to solve
the OLNE. After we use $\tau=1-\exp(-\lambda t)$ to normalize the
infinite time horizon to $[0,1)$, $t$ can be represented by a function
of $\tau$, denoted $t(\tau)=-\ln(1-\tau)/\lambda$, then $P(t)$
and $L(t)$ become functions of $\tau$, i.e., $P(t(\tau))$ and $L(t(\tau))$.
For convenience, we denote them as $P(\tau)$ and $L(\tau)$ respectively.
Thus, the derivative of $P$ over $\tau$ is $P'(\tau)=\dot{P}(t)\times(dt/d\tau)$
and the derivative of $L$ over $\tau$ is $L'(\tau)=\dot{L}(t)\times(dt/d\tau)$.
Since $dt/d\tau=1/(\lambda(1-\tau))$, the ordinary differential equations
(\ref{eq:OLNE-ODE-1D-P})-(\ref{eq:OLNE-ODE-1D_L}) are transformed
to 
\begin{eqnarray}
P'(\tau) & = & \frac{L+f(P)}{\lambda(1-\tau)}\label{eq:OLNE-ODE-1D-transform-P}\\
L'(\tau) & = & \frac{\left[f'(P)-\rho\right]L+\frac{2cP}{n}L^{2}}{\lambda(1-\tau)}\label{eq:OLNE-ODE-1D-transform-L}
\end{eqnarray}
Now we apply the bvp4c solver in Matlab to solve the new boundary
value problem (\ref{eq:OLNE-ODE-1D-transform-P})-(\ref{eq:OLNE-ODE-1D-transform-L}),
while the initial condition is $P(0)=P_{0}$ for one starting state
$P_{0}$ and the terminal condition is set to be $L(\mathscr{T})=L_{ss}$
for a pre-computed total loadings $L_{ss}$ associated with a steady
state $P_{ss}$, where $\mathscr{T}<1$ is chosen to be close to 1.
The initial guess for the solution at one starting point uses the
steady state values $(P_{ss},L_{ss})$ or the solution of the previous
starting point. It is regarded to be successful in finding a solution
only if the terminal $P(\mathscr{T})$ is close to $P_{ss}$. When
multiple steady states exist, the ordinary differential equations
(\ref{eq:OLNE-ODE-1D-transform-P})-(\ref{eq:OLNE-ODE-1D-transform-L})
may admit multiple solutions. In such cases, we select the solution
that yields the highest welfare. Alternatively, it may occur that
only one steady state admits an associated solution, while the system
fails to find solutions corresponding to the others. In cases of non-convergence,
we verify this outcome by experimenting with different values of $\lambda$
in the algorithm and by varying the initial guesses for the bvp4c
solver in MATLAB.

\begin{figure}
\begin{centering}
\includegraphics[width=1\textwidth]{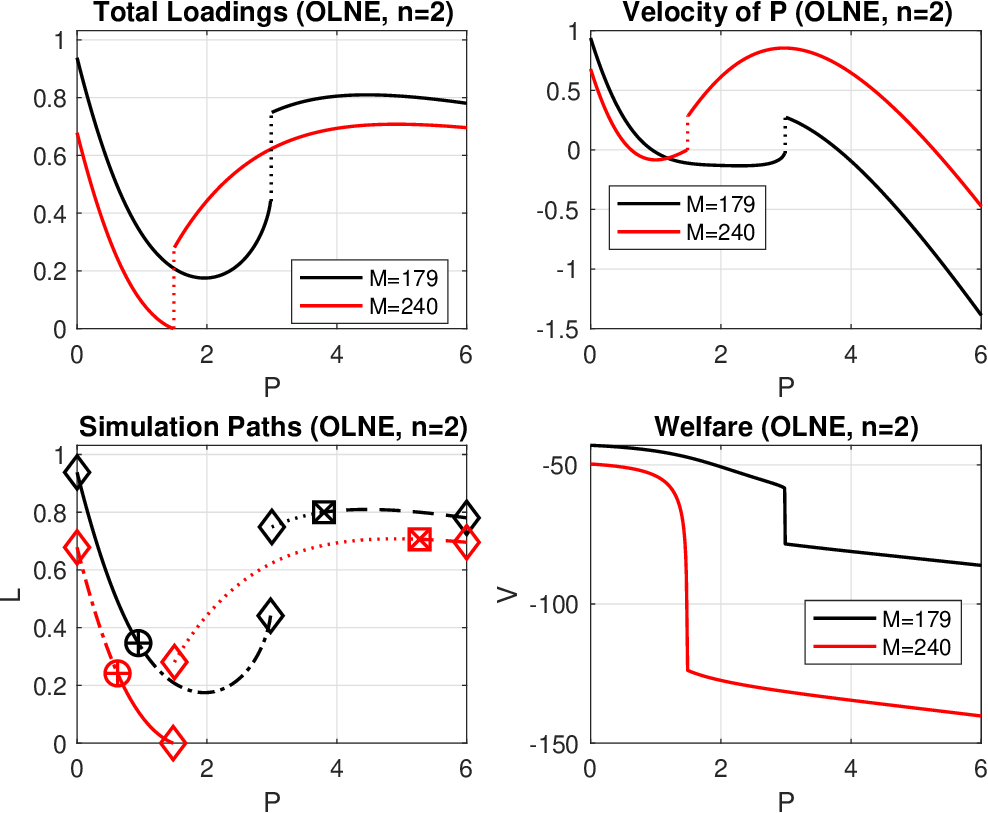}
\par\end{centering}
\caption{\label{fig:OLNE-n2-1D}Solution of One-dimensional Open-loop Nash
Equilibrium when $n=2$}
\end{figure}

For any initial state $P_{0}$, we can apply the above method to solve
the ordinary differential equations (\ref{eq:OLNE-ODE-1D-P})-(\ref{eq:OLNE-ODE-1D_L})
to obtain its associated initial total loadings $L_{0}$. Thus, after
sweeping over a set of initial states $\{P_{0,i}:i=1,...,N\}$ in
the state space, we can obtain their initial total loadings $\{L_{0,i}:i=1,...,N\}$.
That is, we can use the pairs $\{(P_{0,i},L_{0,i}):i=1,...,N\}$ and
piecewise linear interpolation to construct the OLNE strategy function
on the state space at the initial time. Since our optimization problem
is autonomous, it is also the OLNE strategy function at any time.
Figure \ref{fig:OLNE-n2-1D} shows the total loadings $L$ (top-left
panel) and velocity of $P$ (top-right panel), and welfare (bottom-right
panel) as functions of the state variable $P$, as well as simulation
paths (bottom-left panel), for two constant $M=179$, 240, respectively,
with $n=2$ agents. Each strategy function has one jump at a Skiba
point, which is also an unstable steady state. The simulation paths
start with different $P_{0}$ marked with diamond, and converge to
two stable steady states represented with circle, plus, mark, or square,
respectively. Figure \ref{fig:OLNE-n3-1D} in Appendix A displays
a similar picture for $n=3$ agents.

\subsection{Feedback Nash Equilibrium\label{subsec:Feedback-Nash-Equilibrium-1D}}

Let $G(P)$ be the strategy function of one agent under the feedback
Nash equilibrium (FBNE) with $n$ homogeneous agents, where one agent
$a$ has the utility function $\ln(L_{a}(t))-cP(t)^{2}$ at time $t$.
We are solving
\begin{equation}
V(P_{0})=\max_{L_{a}(\cdot)}\int^{\infty}_{0}e^{-\rho t}\left[\ln(L_{a}(t))-cP(t)^{2}\right]dt\label{eq:FBNE-model-1D}
\end{equation}
subject to the transition law (\ref{eq:trans-law-P-1D}), where 
\[
L(t)=L_{a}(t)+(n-1)G(P(t))
\]

From the model (\ref{eq:FBNE-model-1D}), we derive the following
Hamilton-Jacobi-Bellman equation

\begin{equation}
\rho V\left(P\right)=\max_{L_{a}}\left\{ \ln L_{a}-cP^{2}+V'\left(P\right)[L_{a}+\left(n-1\right)G\left(P\right)+f\left(P\right)]\right\} \label{eq:FBNE-HJB-1D}
\end{equation}
Its optimality condition is 
\begin{equation}
\frac{1}{L_{a}}=-V'\left(P\right)\Rightarrow L_{a}=G\left(P\right)=\frac{-1}{V'\left(P\right)},\quad V'\left(P\right)<0\label{eq:FBNE1D-opt-L-cond}
\end{equation}

Substituting the optimality condition into (\ref{eq:FBNE-HJB-1D}),
we obtain the following ordinary differential equation:
\begin{equation}
\rho V\left(P\right)=\ln\left(\frac{-1}{V'\left(P\right)}\right)-cP^{2}-n+V'\left(P\right)f\left(P\right)\label{eq:FBNE-PDE-1D}
\end{equation}

\subsubsection{SFVF Method \label{subsec:SFVF-1D}}

We apply Algorithm 1 to solve the FBNE model (\ref{eq:FBNE-model-1D}).
We choose a set of $N$ equally spaced nodes $\{P_{i}:i=1,...,N\}$
on a pre-specified state space $[P_{\min},P_{\max}]$, where $P_{1}=P_{\min}$
and $P_{N}=P_{\max}$, and the step size is $\Delta=(P_{\max}-P_{\min})/(N-1)$.
\footnote{$N$ is chosen such that a larger number has little impact on the
results.} The specific algorithm to solve the FBNE model (\ref{eq:FBNE-model-1D})
is as follows:
\begin{description}
\item [{Algorithm}] 1.1. SFVF Iteration for FBNE of One-Dimensional Lake
Problem
\item [{Step}] 1. Set an initial guess of the strategy function $G^{0}(P)$
and its associated value function $V^{0}(P)$. Iterate through steps
2-4 for $j=0,1,2,...$, until convergence.
\item [{Step}] 2. Update the strategy function. Solve the following equation
\begin{equation}
\ln\left(x_{i}\right)-cP^{2}_{i}-n-\frac{f(P_{i})}{x_{i}}-\rho V^{j}(P_{i})=0\label{eq:SFI-eq-1D}
\end{equation}
to find $x_{i}$ for each $i=1,...,N$. When multiple solutions for
$x_{i}$ correspond to a given pair $(P_{i},V^{j}(P_{i}))$, proceed
as follows. If the difference between $V^{j}$ and $V^{j-1}$, or
between $G^{j}$ and $G^{j-1}$, is not small, select the solution
that is closest to $G^{j-1}(P_{i})$. Otherwise, choose the solution
for which $-1/x_{i}$ is closest to the derivative of $V^{j}$ at
$P_{i}$, where the derivative can be estimated using a finite difference
method. Use piecewise linear interpolation to construct a loading
strategy function $G^{j+1}(P)$ such that $G^{j+1}(P_{i})=x_{i}$
for all $i$.
\item [{Step}] 3. Update the value function. Use $G^{j+1}(P)$ to generate
a trajectory $(P(t),L_{a}(t))$ starting at $P(0)=P_{i}$ by letting
\[
L_{a}(t)=G^{j+1}(P(t))
\]
and 
\[
P(t+h)=P(t)+\left[nL_{a}(t)+f(P(t))\right]h
\]
where $h$ is a small time step size, and use a numerical quadrature
method to estimate the welfare at $P_{i}$: 
\[
v_{i}=\int^{\infty}_{0}e^{-\rho t}\left[\ln(L_{a}(t))-c\left(P(t)\right)^{2}\right]dt,
\]
and use piecewise linear interpolation to construct a value function
$V^{j+1}(P)$ such that 
\[
V^{j+1}(P_{i})=\omega V^{j}(P_{i})+(1-\omega)v_{i}
\]
with $\omega=0.5$, for each $i=1,...,N$.
\item [{Step}] 4. Check if $V^{j+1}\approx V^{j}$ and $G^{j+1}\approx G^{j}$.
If so, stop the iteration,\footnote{It is also necessary to verify whether the converged value and strategy
functions satisfy the equality of equation (\ref{eq:SFI-eq-1D}).} otherwise go to Step 2 by increasing $j$ with 1.
\end{description}
Algorithm 1.1 might not converge if we do not give a good initial
guess for $V^{0}(P)$ and $G^{0}(P)$. Here we choose the initial
guess in Step 1 of Algorithm 1.1 by the following method: at first,
compute
\[
v_{0,i}=\frac{\log\left(\max\left\{ \frac{-f(P_{i})}{n},0.001\right\} \right)-cP^{2}_{i}}{\rho}
\]
by assuming $P_{i}$ is a steady state, for each $i=1,...,N$. Since
$v_{0,1}$ is negative with large magnitude, $v_{0,i}$ is not decreasing
over $i$, which is inconsistent with that $V'(P)$ is negative. Thus
we construct a series of $\widetilde{v}_{0,i}$ in a backward manner
such that $\widetilde{v}_{0,i}$ is decreasing over $i$: let $\widetilde{v}_{0,N}=v_{0,N}$;
for $i=N,N-1,...,2$, if $\widetilde{v}_{0,i}\geq v_{0,i-1}$, then
$\widetilde{v}_{0,i-1}=\widetilde{v}_{0,i}+\xi\Delta$, otherwise
$\widetilde{v}_{0,i-1}=v_{0,i-1}$. Here we choose $\xi=0.1$. We
then use piecewise linear interpolation to construct $G^{0}(P)$ such
that 
\[
G^{0}(P_{i})=\begin{cases}
\frac{-\Delta}{\widetilde{v}_{0,i+1}-\widetilde{v}_{0,i}}, & i=1,\\
\frac{-2\Delta}{\widetilde{v}_{0,i+1}-\widetilde{v}_{0,i-1}}, & 1<i<N,\\
\frac{-\Delta}{\widetilde{v}_{0,i}-\widetilde{v}_{0,i-1}}, & i=N,
\end{cases}
\]
for $1<i<n$, following finite difference methods to estimate the
derivatives $V'(P_{i})$. With the same method in Step 3 of Algorithm
1.1, we use $G^{0}(P)$ to estimate the welfare $v_{i}$ at $P_{i}$,
then apply piecewise linear interpolation to construct the initial
guess of the value function, $V^{0}(P)$, such that $V^{0}(P_{i})=v_{i}+1$
for all $i$, where the constant 1 is added to make the solution more
accurate around the steady states.

\subsubsection{Results}

\begin{figure}
\begin{centering}
\includegraphics[width=1\textwidth]{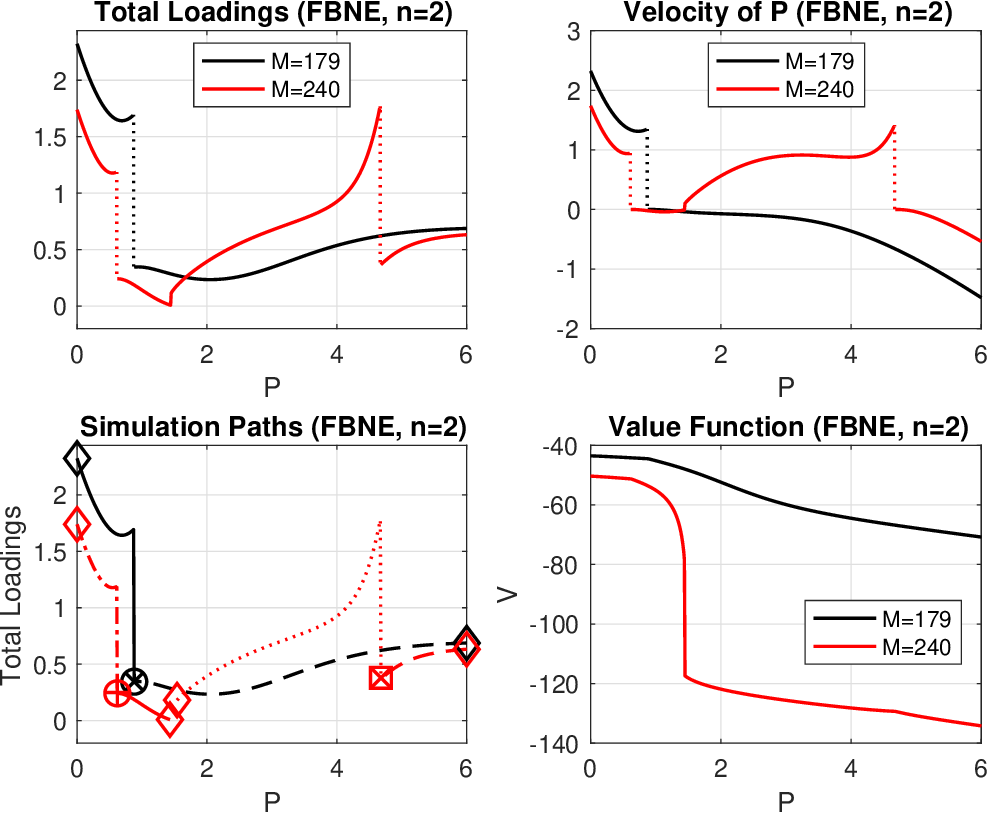}
\par\end{centering}
\caption{\label{fig:FBNE-n2-1D}Solution of One-dimensional Feedback Nash Equilibrium
when $n=2$}
\end{figure}

We apply Algorithm 1.1 to solve the cases with $n=2$ and 3, and $M=179$
and 240, respectively, on the state space $[0,6]$ with $\Delta=0.01$
(i.e., the number of $P_{i}$ is $N=601$). Figure \ref{fig:FBNE-n2-1D}
shows the total loadings $L$ (top-left panel) and velocity of $P$
(top-right panel) at each state $P$, simulation paths of loadings
(bottom-left panel), and the value functions (bottom-right panel),
for $n=2$. The case with $M=179$ has only one steady state, 0.88,
and the case with $M=240$ has three steady states: 0.62, 1.44, and
4.68, among which the middle point (i.e., the Skiba point) is not
stable and the others are stable. These are shown clearly in the simulation
paths starting with different $P_{0}$ marked with diamond, which
converge to stable steady states represented with circle, plus, mark,
or square. Note that the value function has a kink or steep gradient
at a steady state, but it is continuous. When a starting point $P_{0}$
is smaller than the Skiba point, its trajectory converges to the oligotrophic
steady state, but when a starting point $P_{0}$ is larger than the
Skiba point, its trajectory converges to the eutrophic steady state.
Figure \ref{fig:FBNE-n3-1D} in Appendix A for the $n=3$ case has
the same patterns, while the case $n=3$ has higher loadings and lower
welfares than the case $n=2$. Figure \ref{fig:Accuracy-FBNE1D} in
Appendix A displays the values of common logarithms of the difference
between the converged strategy function $G(P)$ and $-1/V'(P)$ at
100 randomly chosen states, where $V(P)$ is the converged value function
and $V'(P)$ is estimated with a finite difference method. It shows
that the difference is small at states that are not close to the locations
where the strategy function jumps. At the jumps, however, $V'$ does
not exist, which leads to larger differences in their vicinity. Compared
with the OLNE solution in Section \ref{subsec:Open-loop-Nash-Equilibrium-1D},
the FBNE solution has different numbers of steady states and jumps
of the strategy functions. Compared with the partial solution provided
by Kossioris et al. (2008), our FBNE solution is defined on the whole
state space $[0,6]$. Moreover, our solution also confirms the finding
of Dockner and Wagener (2014) about the existence of a discontinuous
Markov-perfect equilibrium strategy.

\subsection{Cooperative Solution}

\begin{figure}
\begin{centering}
\includegraphics[width=1\textwidth]{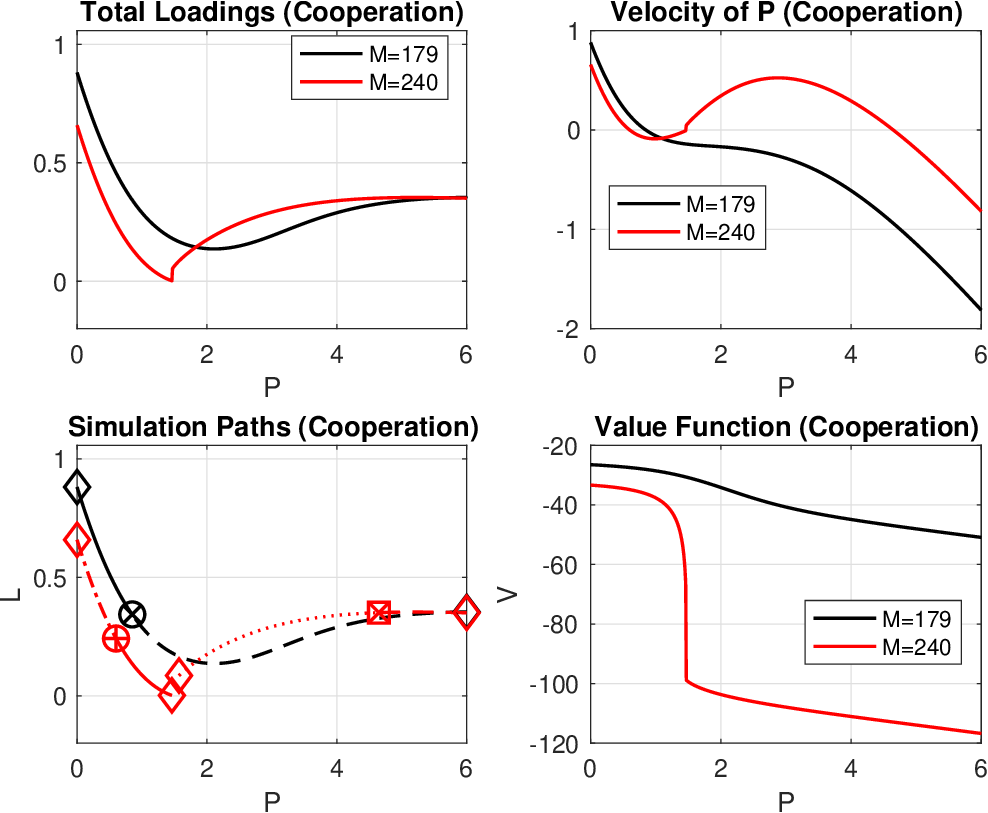}
\par\end{centering}
\caption{\label{fig:opt-1D}Solution of One-dimensional Cooperation}
\end{figure}

We are solving the cooperative problem (\ref{eq:CoopObj}) subject
to the transition law (\ref{eq:tran-law-1D-coop}). We apply Algorithm
1.1 with $n=1$ to solve the problem's cases $M=179$ and 240, respectively,
on the domain $[0,6]$ with $\Delta=0.01$. Since the problem is independent
of the number of agents, the solution paths for $L(t),P(t)$ for the
optimal management with $n=1$ and the cooperative solution with more
than one agent coincide. Aggregate welfare however is different for
different values of $n$, so individual welfare is not proportional
to $n$. Individual welfare for the cooperative solution, as determined
by the value of the system, is presented in Section \ref{sec:Summary-and-Discussion}
for $n=2,3.$ Figure \ref{fig:opt-1D} show the total loadings $L$
(top-left panel) and velocity of $P$ (top-right panel) at each state
$P$, simulation paths of loadings (bottom-left panel), and the value
functions (bottom-right panel). The case with $M=179$ has only one
steady state, 0.85. The case with $M=240$ has three steady states:
0.60, 1.46, and 4.65, among which the middle point (i.e., the Skiba
point) is not stable and the others are stable. These are shown clearly
in the simulation paths starting with different $P_{0}$ marked with
diamond, which converge to stable steady states represented with circle,
plus, mark, or square. Similar to the accuracy check for the FBNE
solutions in Section \ref{subsec:Feedback-Nash-Equilibrium-1D}, Figure
\ref{fig:Accuracy-COOP1D} in Appendix A shows that our cooperative
solution is accurate. Moreover, we also apply our algorithm for solving
OLNE to the cooperative model with only one agent. We find that its
solution is very close to our solution from the SFVF algorithm. Compared
with the FBNE simulation paths in Figure \ref{fig:FBNE-n2-1D}, every
cooperative simulation path is continuous, similar to the OLNE simulation
paths shown in Figure \ref{fig:OLNE-n2-1D}. In contrast, the FBNE
simulation paths that start from the left side of the oligotrophic
steady state for both $M=179$ and 240, or from the left side of the
eutrophic steady state for $M=240$, exhibit jumps around the steady
state to which they converge.

\section{Two-Dimensional Lake Problems}

\subsection{Open-loop Nash Equilibrium\label{subsec:Open-loop-Nash-Equilibrium-2D}}

The two-dimensional OLNE problem solves

\begin{equation}
\max_{L_{a}(\cdot)}\int^{\infty}_{0}e^{-\rho t}\left[\ln(L_{a}(t))-cP(t)^{2}\right]dt\label{eq:open-loop}
\end{equation}
subject to (\ref{eq:trans-law-2D}), where 
\[
L(t)=L_{a}(t)+\sum_{a'\neq a}L_{a'}(t).
\]
It can be transformed to the following ordinary differential equations:
\begin{eqnarray}
\dot{P} & = & L+f(P,M)\label{eq:ODE-system-P}\\
\dot{M} & = & g(P,M)\label{eq:ODE-system-M}\\
\dot{L} & = & \left[f_{P}(P,M)-\rho\right]L+\left[\frac{2cP}{n}-\mu g_{P}(P,M)\right]L^{2}\label{eq:ODE-system-L}\\
\dot{\mu} & = & \left(\rho-g_{M}(P,M)\right)\mu+\frac{f_{M}(P,M)}{L}\label{eq:ODE-system-mu}
\end{eqnarray}

We apply the method in Section \ref{subsec:Method-for-OLNE} to solve
the ordinary differential equations. That is, with $\tau=1-\exp(-\lambda t)$
and its inverse function $t(\tau)=-\ln(1-\tau)/\lambda$, we denote
$P(t(\tau))$, $M(t(\tau))$, $L(t(\tau))$, and $\mu(t(\tau))$ by
$P(\tau)$, $M(\tau)$, $L(\tau)$, and $\mu(\tau)$, respectively,
for convenience. Thus, the derivative of $P$ over $\tau$ is $P'(\tau)=\dot{P}(t)\times(dt/d\tau)$,
and similarly for the other variables. Since $dt/d\tau=1/(\lambda(1-\tau))$,
the ordinary differential equations (\ref{eq:ODE-system-P})-(\ref{eq:ODE-system-mu})
are transformed to 
\begin{eqnarray}
P'(\tau) & = & \frac{L+f(P,M)}{\lambda(1-\tau)}\label{eq:ODE-system-P-1}\\
M'(\tau) & = & \frac{g(P,M)}{\lambda(1-\tau)}\label{eq:ODE-system-M-1}\\
L'(\tau) & = & \frac{\left[f_{P}(P,M)-\rho\right]L+\left[\frac{2cP}{n}-g_{P}(P,M)\mu\right]L^{2}}{\lambda(1-\tau)}\label{eq:ODE-system-L-1}\\
\mu'(\tau) & = & \frac{\left(\rho-g_{M}(P,M)\right)\mu L+f_{M}(P,M)}{\lambda(1-\tau)L}\label{eq:ODE-system-mu-1}
\end{eqnarray}
We then apply the bvp4c solver in Matlab to solve the new boundary
value problem (\ref{eq:ODE-system-P-1})-(\ref{eq:ODE-system-mu-1}),
while the initial condition is $(P(0),M(0))=(P_{0},M_{0})$ for one
starting state $(P_{0},M_{0})$ and the terminal condition is set
to be $(L(\mathscr{T}),\mu(\mathscr{T}))=(L_{ss},\mu_{ss})$ for a
pre-computed $(L_{ss},\mu_{ss})$ associated with a steady state $(P_{ss},M_{ss})$,
where $\mathscr{T}<1$ is chosen to be close to 1. The initial guess
for the solution at one starting point uses the steady state values
$(P_{ss},M_{ss},L_{ss},\mu_{ss})$ or the solution of the previous
starting point. It is regarded to be successful in finding a solution
only if the terminal state $(P(\mathscr{T}),M(\mathscr{T}))$ is close
to the steady state $(P_{ss},M_{ss})$. Similar to solving the OLNE
for the one-dimensional lake problems, when multiple steady states
exist, the ordinary differential equations (\ref{eq:ODE-system-P-1})-(\ref{eq:ODE-system-mu-1})
may admit multiple solutions. In such cases, we select the solution
that yields the highest welfare. Alternatively, it may occur that
only one steady state admits an associated solution, while the system
fails to find solutions corresponding to the others.

\begin{figure}
\begin{centering}
\includegraphics[width=1\textwidth]{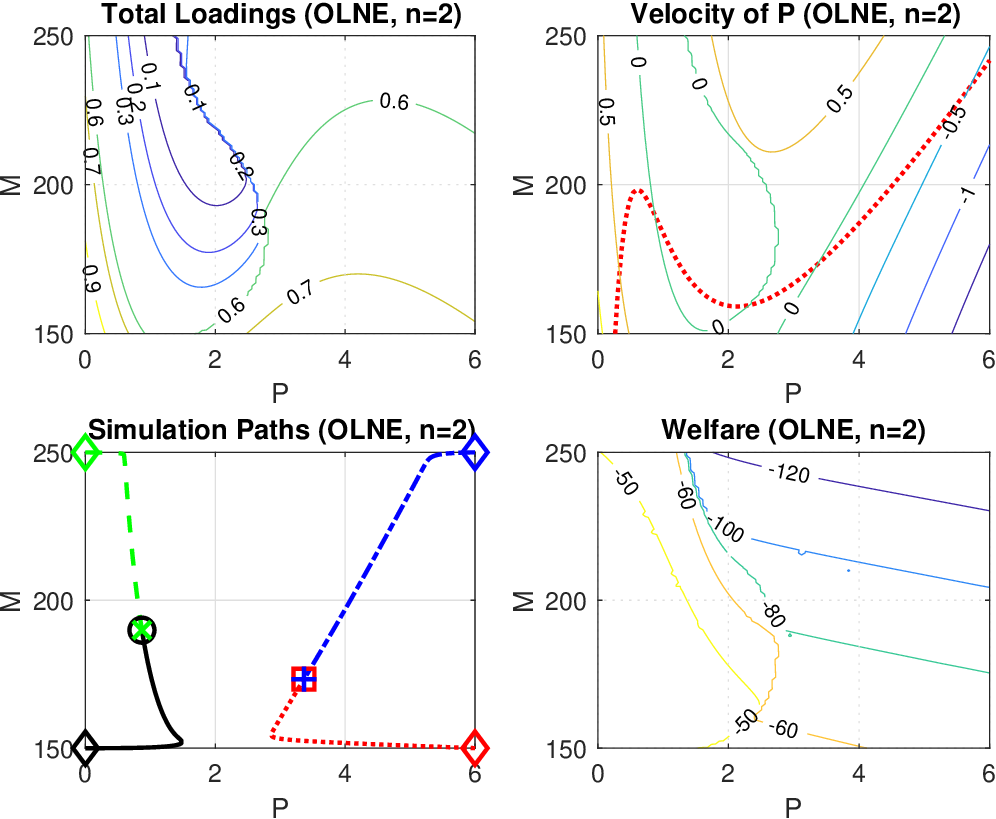}
\par\end{centering}
\caption{\label{fig:OLNE-n2}Solution of Two-dimensional OLNE when $n=2$.
The red dotted line on the top-right panel represents the isocline
$\dot{M}=0$.}
\end{figure}

Figure \ref{fig:OLNE-n2} shows the total loadings $L$ of the model
(\ref{eq:open-loop}) at $(P,M)$, velocity of $P$, four simulation
paths starting from corner points, and welfare for $n=2$. Figure
\ref{fig:OLNE-n2} matches closely with Figure 4 of Grass et al. (2017),
as the full cooperative model in Grass et al. (2017) with $c_{M}=0.0868$
has the same ordinary differential equations with the OLNE model (\ref{eq:open-loop})
with $c=0.1736$, for $n=2$. Moreover, Figure \ref{fig:OLNE-n2}
shows an irreversibility manifold in the loadings (top-left panel).
This manifold is reflected by the dense contour lines of loadings
in the top-left panel, the dense contour lines of welfares in the
bottom-right panel, and the middle isocline $\dot{P}=0$ in the top-right
panel. The loadings are discontinuous across the two sides of the
irreversibility manifold. The velocity of $P$ is negative on the
left of the manifold, or positive on the right of the manifold, so
the irreversibility manifold has similar pattern of a Skiba manifold.
However, welfare is discontinuous across the irreversibility manifold:
the left side of the manifold yields substantially higher welfare
than the right side, while the loadings on the left side are smaller.
Therefore, an agent would choose the smaller loadings if a starting
point is on the irreversibility manifold, while an agent will be indifferent
if a starting point is on a Skiba manifold. Figure \ref{fig:OLNE-n3}
for the $n=3$ case in Appendix A has the same patterns, and it matches
closely with Figure 5 of Grass et al. (2017) with $c_{L}=0.057867$.

\subsection{Feedback Nash Equilibrium\label{subsec:Feedback-Nash-Equilibrium-2D}}

Assume $L_{a}=G\left(P,M\right)$ at a state $(P,M)$ is the loading
strategy of agent $a$ under the feedback Nash equilibrium with $n$
homogenous agents, each of whom having the utility function $\ln(L_{a}(t))-cP(t)^{2}$,
for $a=1,...,n$. The two-dimensional FBNE problem solves

\begin{equation}
V(P_{0},M_{0})=\max_{L_{a}(\cdot)}\int^{\infty}_{0}e^{-\rho t}\left[\ln(L_{a}(t))-cP(t)^{2}\right]dt\label{eq:FBNE-model-2D}
\end{equation}
subject to (\ref{eq:trans-law-2D}) where $(P(0),M(0))=(P_{0},M_{0})$
is an initial state vector, and
\begin{eqnarray*}
L(t) & = & L_{a}(t)+(n-1)G(P(t),M(t)).
\end{eqnarray*}

The value function $V\left(P,M\right)$ satisfies the following Hamilton-Jacobi-Bellman
equation

\begin{eqnarray}
\rho V\left(P,M\right) & = & \max_{L_{a}}\left\{ \ln L_{a}-cP^{2}\right.\nonumber \\
 &  & +V_{P}\left(P,M\right)[L_{a}+\left(n-1\right)G\left(P,M\right)+f\left(P,M\right)]\nonumber \\
 &  & +V_{M}\left(P,M\right)g(P,M)\Bigr\}\label{eq:valuefun_feedback}
\end{eqnarray}
Its optimality condition is 
\begin{equation}
\frac{1}{L_{a}}=-V_{P}\left(P,M\right)\Rightarrow L_{a}=G\left(P,M\right)=\frac{-1}{V_{P}\left(P,M\right)},\quad V_{P}\left(P,M\right)<0
\end{equation}

Substituting the optimality condition into (\ref{eq:valuefun_feedback}),
we obtain 
\begin{eqnarray}
\rho V\left(P,M\right) & = & \ln\left(\frac{-1}{V_{P}\left(P,M\right)}\right)-cP^{2}-n\nonumber \\
 &  & +V_{P}\left(P,M\right)f\left(P,M\right)+V_{M}\left(P,M\right)g\left(P,M\right)\label{eq:FBNE-PDE-2D}
\end{eqnarray}
and 
\begin{equation}
L=nL_{a}=\frac{-n}{V_{P}\left(P,M\right)}\label{eq:FB-L}
\end{equation}

\subsubsection{SFVF Method}

We apply Algorithm 1 to solve the FBNE model (\ref{eq:FBNE-model-2D}).
We choose a tensor grid of $N_{1}\times N_{2}$ nodes $\{(P_{i_{1}},M_{i_{2}}):1\leq i_{1}\leq N_{1},1\leq i_{2}\leq N_{2}\}$
in a pre-specified state space $[P_{\min},P_{\max}]\times[M_{\min},M_{\max}]$,
where $P_{1}=P_{\min}$, $P_{N_{1}}=P_{\max}$, $M_{1}=M_{\min}$,
and $M_{N_{2}}=M_{\max}$, where the step sizes, $\Delta_{1}=(P_{\max}-P_{\min})/(N_{1}-1)$
and $\Delta_{2}=(M_{\max}-M_{\min})/(N_{2}-1)$, are small. The specific
algorithm to solve the FBNE model (\ref{eq:FBNE-model-2D}) is as
follows:
\begin{description}
\item [{Algorithm}] 1.2. SFVF Iteration for FBNE of Two-Dimensional Lake
Problem
\item [{Step}] 1. Set an initial guess of the strategy function, $G^{0}(P,M)$,
and its associated value function $V^{0}(P,M)$. Iterate through steps
2-5 for $j=0,1,2,...$, until convergence.
\item [{Step}] 2. Use $V^{j}(P,M)$ and finite difference methods to estimate
the partial derivative of the value function over $M$: 
\[
V^{j}_{M}(P_{i_{1}},M_{i_{2}})\approx w_{i_{1},i_{2}}=\begin{cases}
\frac{V^{j}(P_{i_{1}},M_{i_{2}+1})-V^{j}(P_{i_{1}},M_{i_{2}})}{\Delta_{2}}, & i_{2}=1,\\
\frac{V^{j}(P_{i_{1}},M_{i_{2}+1})-V^{j}(P_{i_{1}},M_{i_{2}-1})}{2\Delta_{2}}, & 1<i_{2}<N_{2},\\
\frac{V^{j}(P_{i_{1}},M_{i_{2}})-V^{j}(P_{i_{1}},M_{i_{2}-1})}{\Delta_{2}}, & i_{2}=N_{2},
\end{cases}
\]
for each $i_{1}$ and $i_{2}$.
\item [{Step}] 3. Update the strategy function. Solve the following equation
\begin{equation}
\ln\left(x_{i_{1},i_{2}}\right)-cP^{2}_{i_{1}}-n-\frac{f(P_{i_{1}},M_{i_{2}})}{x_{i_{1},i_{2}}}+w_{i_{1},i_{2}}g\left(P_{i_{1}},M_{i_{2}}\right)-\rho V^{j}(P_{i_{1}},M_{i_{2}})=0\label{eq:SFI-eq-2D}
\end{equation}
to find $x_{i_{1},i_{2}}$ for each $i_{1}$ and $i_{2}$. When multiple
solutions for $x_{i_{1},i_{2}}$ correspond to a given $(P_{i_{1}},M_{i_{2}},V^{j}(P_{i_{1}},M_{i_{2}}))$,
proceed as follows. If the difference between $V^{j}$ and $V^{j-1}$,
or between $G^{j}$ and $G^{j-1}$, is not small, select the solution
that is closest to $G^{j-1}(P_{i_{1}},M_{i_{2}})$. Otherwise, choose
the solution for which $-1/x_{i_{1},i_{2}}$ is closest to $V^{j}_{P}(P_{i_{1}},M_{i_{2}})$,
which can be estimated using a finite difference method. Use piecewise
bilinear interpolation to construct a loading strategy function $G^{j+1}(P,M)$
such that $G^{j+1}(P_{i_{1}},M_{i_{2}})=x_{i_{1},i_{2}}$ for all
$i_{1}$ and $i_{2}$.
\item [{Step}] 4. Update the value function. Use $G^{j+1}(P,M)$ to generate
a trajectory $(P(t),M(t),L_{a}(t))$ starting at $P(0)=P_{i_{1}}$
and $M(0)=M_{i_{2}}$ by letting 
\[
L_{a}(t)=G^{j+1}(P(t),M(t))
\]
and 
\begin{eqnarray*}
P(t+h) & = & P(t)+\left[nL_{a}(t)+f(P(t),M(t))\right]h\\
M(t+h) & = & M(t)+g(P(t),M(t))h
\end{eqnarray*}
where $h$ is a small time step size, and compute 
\[
v_{i_{1},i_{2}}=\int^{\infty}_{0}e^{-\rho t}\left[\ln(L_{a}(t))-cP(t)^{2}\right]dt,
\]
and use piecewise bilinear interpolation to construct a value function
$V^{j+1}(P,M)$ such that 
\[
V^{j+1}(P_{i_{1}},M_{i_{2}})=\omega V^{j}(P_{i_{1}},M_{i_{2}})+(1-\omega)v_{i_{1},i_{2}}
\]
with $\omega=0.5$, for all $i_{1}$ and $i_{2}$.
\item [{Step}] 5. Check if $V^{j+1}\approx V^{j}$ and $G^{j+1}\approx G^{j}$.
If so, stop the iteration, otherwise go to Step 2 by increasing $j$
with 1.
\end{description}
Compared with Algorithm 1.1, Algorithm 1.2 has one more Step 2 for
estimating the partial derivative of the value function over $M$.
Since Algorithm 1.1 has only one state variable $P$, this step is
not needed. In fact the combination of Steps 2 and 3 in Algorithm
1.2 is equivalent to Step 2 of Algorithm 1. Because the value function
exhibits steep gradients in the neighborhood of the weak Skiba manifold,
denoted $\Omega$, numerical errors in estimating $V^{j}_{M}(P_{i_{1}},M_{i_{2}})$
in Step 2 in Algorithm 1.2 can be large if $\Delta_{2}$ is not sufficiently
small. Consequently, equation (\ref{eq:SFI-eq-2D}) in Step 3 in Algorithm
1.2 may have no solution; in that case, the algorithm sets $x_{i_{1},i_{2}}=G^{j}(P_{i_{1}},M_{i_{2}})$.
Even when a solution exists, it may deviate from the true solution
due to these numerical inaccuracies. However, due to the curse-of-dimensionality
of the tensor grid $(P_{i_{1}},M_{i_{2}})$ on the state space, it
is time-consuming to run the algorithm with a sufficiently small $\Delta_{2}$.
Fortunately, since no trajectory can come from outside of $\Omega$
to inside of $\Omega$, the inaccurate solution in $\Omega$ has no
impact on the solution outside of $\Omega$, that is, the solution
outside of $\Omega$ can still be accurate.

Algorithm 1.2 might not converge if we do not give a good initial
guess for $V^{0}(P,M)$ and $G^{0}(P,M)$. Here we choose an initial
guess by the method similar to that in Algorithm 1.1: at first, compute
\[
v_{0,i_{1},i_{2}}=\frac{\log\left(\max\left\{ \frac{-f(P_{i_{1}},M_{i_{2}})}{n},0.01\right\} \right)-cP^{2}_{i_{1}}}{\rho}
\]
for each $i_{1}$ and $i_{2}$. Next we construct a series of $\widetilde{v}_{0,i_{1},i_{2}}$
in a backward manner such that $\widetilde{v}_{0,i_{1},i_{2}}$ is
decreasing over $i_{1}$ and $i_{2}$, like what we do in Section
\ref{subsec:SFVF-1D}. We then use piecewise bilinear interpolation
to construct $G^{0}(P,M)$ such that 
\[
G^{0}(P_{i_{1}},M_{i_{2}})=\begin{cases}
\frac{-\Delta_{1}}{\widetilde{v}_{0,i_{1}+1,i_{2}}-\widetilde{v}_{0,i_{1},i_{2}}}, & i_{1}=1,\\
\frac{-2\Delta_{1}}{\widetilde{v}_{0,i_{1}+1,i_{2}}-\widetilde{v}_{0,i_{1}-1,i_{2}}}, & 1<i_{1}<N_{1},\\
\frac{-\Delta_{1}}{\widetilde{v}_{0,i_{1},i_{2}}-\widetilde{v}_{0,i_{1}-1,i_{2}}}, & i_{1}=N_{1},
\end{cases}
\]
for all $i_{1}$ and $i_{2}$. Using the same method as in Step 4
of Algorithm 1.2, we use $G^{0}(P,M)$ to estimate the welfare $v_{i_{1},i_{2}}$
at $(P_{i_{1}},M_{i_{2}})$, and set $V^{0}(P_{i_{1}},M_{i_{2}})=v_{i_{1},i_{2}}-\max_{i_{1},i_{2}}\left\{ v_{i_{1},i_{2}}\right\} /2$
for all $i_{1}$ and $i_{2}$. The constant term $\max_{i_{1},i_{2}}\left\{ v_{i_{1},i_{2}}\right\} /2$
reduces the likelihood that equation (\ref{eq:SFI-eq-2D}) has no
solution.

\subsubsection{Results}

\begin{figure}
\begin{centering}
\includegraphics[width=1\textwidth]{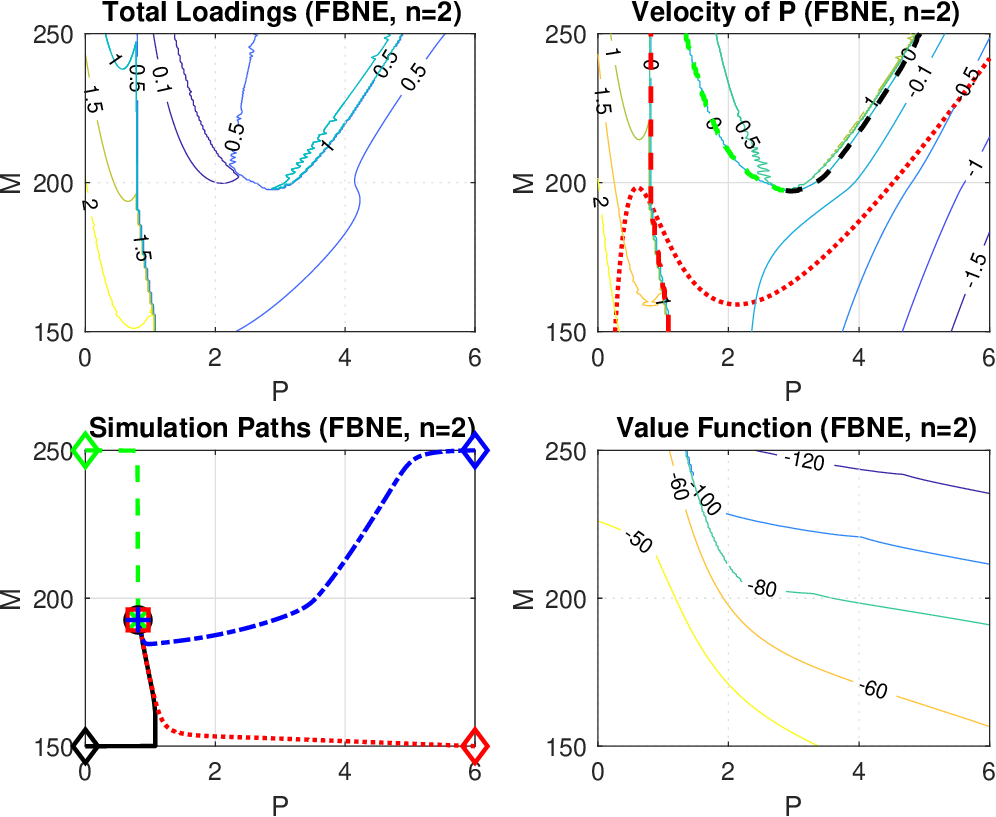}
\par\end{centering}
\caption{\label{fig:FBNE-n2}Solution of Two-dimensional Feedback Nash Equilibrium
when $n=2$. The red dotted line on the top-right panel represents
the isocline $\dot{M}=0$.}
\end{figure}

We apply Algorithm 1.2 to solve the FBNE model (\ref{eq:FBNE-model-2D})
on the approximation domain $[0,6]\times[150,200]$, for $n=2$ or
3. We choose $201\times201$ tensor grids $(P_{i_{1}},M_{i_{2}})$
on the domain. Figure \ref{fig:FBNE-n2} shows the total loadings
$L$ at states $(P,M)$, velocity of $P$ at states $(P,M)$, four
simulation paths starting from corner points, and the value function,\footnote{The velocity of $M$ at states is independent of decisions, so we
do not plot it except the isocline line $\dot{M}=0$ in the top-right
panel for velocity of $P$.} for $n=2$. The small wiggles on the isoclines on the top-left panel
and the top-right panel of Figure \ref{fig:FBNE-n2} are caused by
numerical errors.

The top-right panel of Figure \ref{fig:FBNE-n2} shows the contours
of velocity of $P$ and the isocline $\dot{M}=0$. It shows that there
are three isoclines $\dot{P}=0$ highlighted by dashed lines. The
red dashed isocline $\dot{P}=0$ and the isocline $\dot{M}=0$ (the
red dotted line on the top-right panel) have one intersection point
$(0.81,193)$, i.e., the steady state. It is also the converged point
of the four simulation paths. Moreover, the velocity of $M$ is negative
above the isocline $\dot{M}=0$ and positive below the isocline $\dot{M}=0$
in the approximation domain $[0,6]\times[150,200]$, and the contour
of velocity of $P$ shows that $\dot{P}$ decreases over $P$ from
positive values to negative values (i.e., if $P$ is in the left part
of the red dashed isocline $\dot{P}=0$, then its next step will be
to move right, and if $P$ is in the right part of the red dashed
isocline $\dot{P}=0$, then its next step will be to move left). Thus,
we see that the steady state is stable. This is also shown from the
four simulation paths starting from corner points in the bottom-left
panel of Figure \ref{fig:FBNE-n2}: each converges to the steady state
$(0.81,193)$.

The green dashed isocline $\dot{P}=0$ is the weak Skiba manifold,
corresponding to the Skiba point in the one-dimensional FBNE solution,
and the black dashed isocline $\dot{P}=0$ corresponds to the eutrophic
steady state in the one-dimensional FBNE solution with the large constant
$M=240$. The weak Skiba manifold is also reflected by steep gradients
of the value function on the same locations of $(P,M)$, as shown
on the bottom-right panel of Figure \ref{fig:FBNE-n2}, and this is
also consistent with the pattern of the value function of the one-dimensional
FBNE in Figure \ref{fig:FBNE-n2-1D}. Figure \ref{fig:FBNE-n3} in
Appendix A for the case $n=3$ looks similar to Figure \ref{fig:FBNE-n2},
except that the total loadings are higher in general, the value function's
values are smaller, and the steady state becomes $(0.87,190)$.\footnote{The FBNE solution with $n=4$ is also similar, so we omit it.}
Figure \ref{fig:Accuracy-FBNE2D} in Appendix A displays the contour
of common logarithms of the difference between the converged strategy
function $G(P,M)$ and $-1/V_{P}(P,M)$ at 10,000 randomly chosen
states for $n=2$ or 3, where $V(P,M)$ is the converged value function
and $V_{P}(P,M)$ is estimated with a finite difference method. It
shows that the difference is small at states that are not close to
the locations where the strategy function jumps (i.e., along the dashed
lines). At the jumps, however, neither $V_{P}(P,M)$ nor $V_{M}(P,M)$
exists in theory. As a result, the numerical estimation of $V_{P}(P,M)$
used in computing the difference, as well as that of $V_{M}(P,M)$
in Step 2 of Algorithm 1.2, becomes inaccurate, leading to larger
differences in their vicinity. Compared to the red dashed line, the
black dashed line corresponds to a larger magnitude of $g(P,M)$,
which amplifies the impact of inaccuracies in estimating $V_{M}(P,M)$
during Step 3 of Algorithm 1.2. This explains why the accuracy around
the black dashed line is lower than around the red dashed line. The
reduced accuracy near the green dashed line (i.e., the weak Skiba
manifold) is attributable to large numerical errors in estimating
$V_{M}(P,M)$, as the value function exhibits steep gradients around
the manifold. When a state point lies above the green and black dashed
lines, the difference becomes relatively large\textemdash often limited
to two-digit accuracy. Moreover, the closer the point is to the black
dashed line, the greater the difference tends to be. This behavior
arises because a trajectory originating from such a point will cross
the black dashed line, where the strategy function exhibits discontinuities.
These jumps reduce the accuracy of computing $V$ at the initial state
when it is evaluated along the trajectory using piecewise bilinear
interpolation of the strategy function in Step 4 of Algorithm 1.2.
The effect is more pronounced for points closer to the black dashed
line due to the influence of the discounting rate $\rho>0$. 

\subsection{Cooperative Solution}

\begin{figure}
\begin{centering}
\includegraphics[width=1\textwidth]{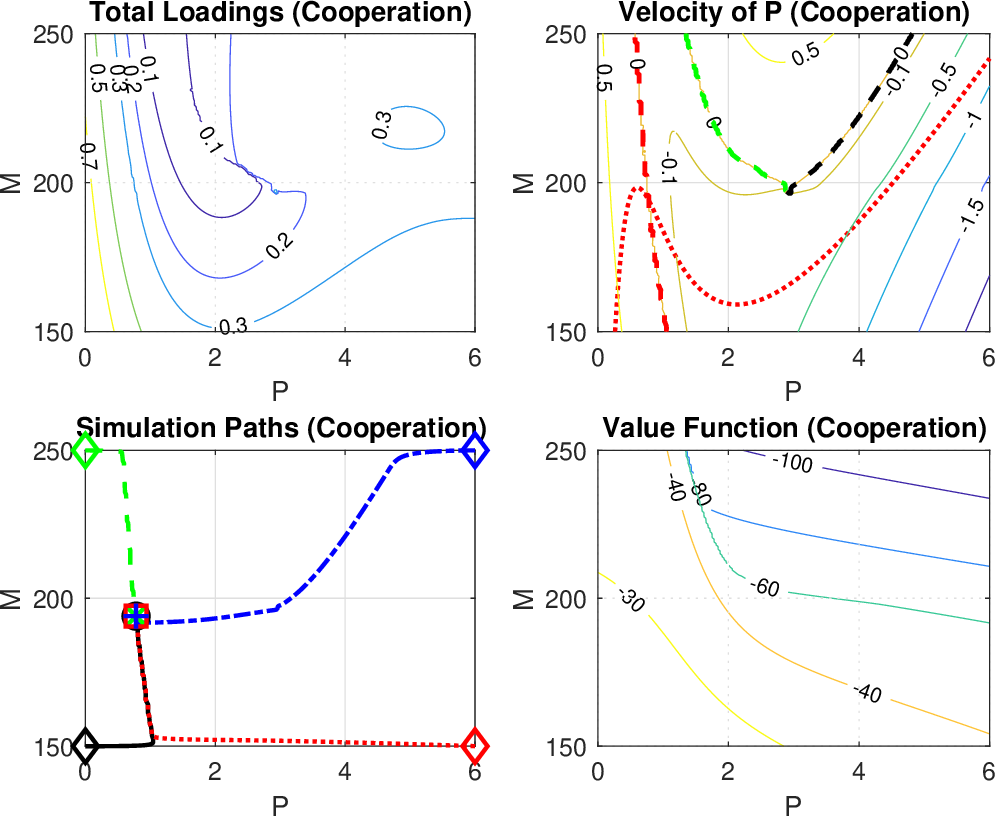}
\par\end{centering}
\caption{\label{fig:opt-twoD}Solution of Two-dimensional Cooperation. The
red dotted line on the top-right panel represents the isocline $\dot{M}=0$.}
\end{figure}

The two-dimensional cooperative problem solves

\[
V(P_{0},M_{0})=\max_{L(\cdot)}\int^{\infty}_{0}e^{-\rho t}\left[\ln(L(t))-cP(t)^{2}\right]dt
\]
subject to (\ref{eq:trans-law-2D}) where $(P(0),M(0))=(P_{0},M_{0})$
is an initial state vector. Similar to solving the FBNE, we apply
Algorithm 1.2 with $n=1$ to solve the problem on the approximation
domain $[0,6]\times[150,200]$. Figure \ref{fig:opt-twoD} shows the
total loadings $L$ at states $(P,M)$, velocity of $P$ at states
$(P,M)$, four simulation paths starting from corner points, and the
value function.

The top-right panel of Figure \ref{fig:opt-twoD} shows the contours
of velocity of $P$ and the isocline $\dot{M}=0$. It shows that there
are three isoclines $\dot{P}=0$. The red dashed isocline $\dot{P}=0$
and $\dot{M}=0$ have one intersection point $(0.78,194)$, i.e.,
the steady state. It is also the converged point of the four simulation
paths. Moreover, from the velocity of $M$ and $P$, we see that the
steady state is stable. The green dashed isocline $\dot{P}=0$ is
the weak Skiba manifold, corresponding to the Skiba point in the one-dimensional
solution of the cooperative problem, and the black dashed isocline
$\dot{P}=0$ corresponds to the eutrophic steady state in the one-dimensional
cooperative solution with the large constant $M=240$. The weak Skiba
manifold is also reflected by steep gradients of the value function
on the same locations of $(P,M)$, as shown on the bottom-right panel
of Figure \ref{fig:opt-twoD}, and this is also consistent with the
pattern of the value function of the one-dimensional cooperative model
in Figure \ref{fig:opt-1D}. Our cooperative solution is also consistent
with the cooperative solution of Grass et al. (2017) obtained from
the OCMat toolbox (Grass, 2012). Similar to the accuracy check for
the FBNE solutions in Section \ref{subsec:Feedback-Nash-Equilibrium-2D},
Figure \ref{fig:Accuracy-COOP2D} in Appendix A shows that our cooperative
solution achieves high accuracy across the entire state space, except
in the neighborhood of the weak Skiba manifold. The reduced accuracy
near the weak Skiba manifold is due to large numerical errors in estimating
the partial derivative of the value function over $M$, where the
function exhibits steep gradients around the manifold.

\section{Summary and Discussion\label{sec:Summary-and-Discussion}}

Our results for the cooperative, OLNE and FBNE solutions for the one-dimensional
and two-dimensional lake problems are summarized in table 1, in which
$\left(P^{*},M^{*},L^{*},V^{*}\right)$ are values at steady state
and $V$-range is the range of individual welfare at the entire state
space used for the different solutions. Welfare at the cooperative
solution, given by $V^{*},$ and $V$-range, is adjusted for individuals
with the number of agents.\footnote{An individual welfare with $n$ cooperative agents is the difference
between the welfare of optimal management (with the total number of
agents being one) and $(\ln n)/\rho$, due to the logarithmic form
in the utility function and the assumption of agent symmetry.}

Table 1: Results

\begin{tabular}{c|c|c|c}
\hline 
$1D$ & \multicolumn{3}{c}{$n=2$}\tabularnewline
\hline 
$M_{}=179$ & COOP & \multirow{1}{*}{OLNE} & FBNE\tabularnewline
\hline 
$P^{*}$ & $0.85$ & $0.95,2.98,3.81$ & $0.88$\tabularnewline
$L^{*}$ & $0.34$ & $0.34,0.44,0.8$ & $0.34$\tabularnewline
$V^{*}$ & $-44$ & $-45$, $-58$, $-81$ & $-45$\tabularnewline
$V$-range & $(-43,-67)$ & $(-43,-86)$ & $(-44,-71)$\tabularnewline
\hline 
 & \multicolumn{3}{c}{$n=3$}\tabularnewline
\hline 
$M=179$ & COOP & OLNE & FBNE\tabularnewline
\hline 
$P^{*}$ & $0.85$ & $0.99,2.51,4.56$ & $0.92$\tabularnewline
$L^{*}$ & $0.34$ & $0.35,1.35,1.21$ & $0.35$\tabularnewline
$V^{*}$ & $-54$ & $-55$, $-65$, $-106$ & $-54$\tabularnewline
$V$-range & $(-53,-77)$ & $(-53,-110)$ & $(-54,-86)$\tabularnewline
\hline 
\end{tabular}

\begin{tabular}{c|c|c|c}
\hline 
$1D$ & \multicolumn{3}{c}{$n=2$}\tabularnewline
\hline 
$M=240$ & COOP & \multirow{1}{*}{OLNE} & FBNE\tabularnewline
\hline 
$P^{*}$ & $0.6,1.46,4.65$ & 0.63, 1.48, $5.28$ & 0.62, 1.44, $4.68$\tabularnewline
$L^{*}$ & $0.24,0.002,0.35$ & $0.24,0.0003,0.71$ & $0.24,0.007,0.37$\tabularnewline
$V^{*}$ & $-51,-82,-129$ & $-51,$ $-106$, $-124$ & $-51,$ $-78$, $-129$\tabularnewline
$V$-range & $(-49,-133)$ & $(-50,-140)$ & $(-50,-134)$\tabularnewline
\hline 
 & \multicolumn{3}{c}{$n=3$}\tabularnewline
\hline 
$M_{}=240$ & COOP & OLNE & FBNE\tabularnewline
\hline 
$P^{*}$ & $0.6,1.46,4.65$ & $0.64,1.48,5.80$ & $0.64,1.4,4.7$\tabularnewline
$L^{*}$ & $0.24,0.002,0.35$ & $0.24,0.0003,1.04$ & $0.24,0.02,0.38$\tabularnewline
$V^{*}$ & $-61,-92,-139$ & $-61,$ $-116$, $-162$ & $-61,$ $-85$, $-139$\tabularnewline
$V$-range & $(-59,-143)$ & $(-59,-163)$ & ($-61,-145)$\tabularnewline
\hline 
\end{tabular}

\begin{tabular}{c|c|c|c}
\hline 
$2D$ & \multicolumn{3}{c}{$n=2$}\tabularnewline
\hline 
 & COOP & \multirow{1}{*}{OLNE} & FBNE\tabularnewline
\hline 
$\left(P^{*},M^{*}\right)$ & ($0.78,194$) & $(0.87,190),(2.34,160),(3.37,173)$ & $(0.81,193)$\tabularnewline
$L^{*}$ & 0.31 & 0.32, 0.49, 0.68 & 0.31\tabularnewline
$V^{*}$ & $-46$ & $-45,-48,-71$ & $-46$\tabularnewline
$V$-range & $(-39,-130)$ & $(-40,-137)$ & $(-40,-132)$\tabularnewline
\hline 
 & \multicolumn{3}{c}{$n=3$}\tabularnewline
\hline 
 & COOP & OLNE & FBNE\tabularnewline
\hline 
$\left(P^{*},M^{*}\right)$ & ($0.78,194$) & $(4.81,208)$ & $(0.87,190)$\tabularnewline
$L^{*}$ & 0.31 & 0.93 & 0.32\tabularnewline
$V^{*}$ & $-56$ & $-121$ & $-56$\tabularnewline
$V$-range & $(-49,-140)$ & $(-72,-158)$ & $(-50,-144)$\tabularnewline
\hline 
\end{tabular}

In the one-dimensional formulation of the lake problem under conditions
of low mud stock, the Cooperative (COOP) and the FBNE solutions exhibit
qualitatively similar characteristics. Both approaches converge to
a single oligotrophic, stable steady state, with nearly identical
values for the state and control variables: phosphorus stock $(P)$
and phosphorus loading $(L)$, respectively. Increasing the number
of agents from two to three does not significantly affect the steady-state
levels of $(P,L)$ although it does result in a reduction in welfare.

In contrast, the OLNE solution is characterized by the existence of
three steady states. Among these, the intermediate steady state is
unstable and a Skiba point. Relative to the COOP and FBNE outcomes,
the OLNE solution yields lower welfare. Thus, under low $M$, the
FBNE solution leads to a stable oligotrophic steady state that closely
approximates the cooperative benchmark in both quantitative measures
and welfare outcomes. The OLNE solution, however, allows for convergence
to either an oligotrophic or a eutrophic steady state, depending on
the initial conditions.

When the model is solved for higher values of $M$, all solution types
(COOP, FBNE, and OLNE) yield three steady states, with the intermediate
one being unstable across all solution concepts. Notably, the location
of this unstable steady state is close across the three solution types.
Around the steady states, the FBNE solution remains very similar to
the COOP solution in terms of phosphorus stock, phosphorus loading
values, and welfare. The OLNE solution diverges, exhibiting higher
phosphorus levels at both the oligotrophic and eutrophic steady states.
This general pattern persists for both $n=2$ and $n=3$ agents. At
states not near to the steady states, the FBNE solution has larger
loadings than the COOP solution (by comparing Figures \ref{fig:FBNE-n2-1D}
and \ref{fig:FBNE-n3-1D} to Figure \ref{fig:opt-1D}), implying faster
convergence to the associate steady states, but the FBNE's individual
welfare is only slightly lower than the COOP's individual welfare
according to the $V$-ranges in Table 1, unless the starting state
$P$ is large and the constant $M$ is small.

In the two-dimensional version of the problem, the relationship between
the COOP and FBNE solutions remains consistent with the one-dimensional
case. For $n=2$ or 3, both solutions converge to an oligotrophic
steady state in the $(P,M)$ state space, with similar values for
phosphorus stock, phosphorus loading, mud stock, and welfare. However,
the FBNE solution yields a slightly higher steady-state phosphorus
stock and phosphorus loading compared with the COOP solution. Welfare
levels are comparable between the COOP and FBNE solutions at all states,
while the FBNE solution prescribes higher loadings at states that
are not close to the steady state.

In the case of $n=2$, the OLNE solution produces three steady states
in the $(P,M)$ space, with the oligotrophic steady state exhibiting
higher phosphorus stock levels compared to the corresponding steady
states under the COOP and FBNE solutions. When the number of agents
increases to $n=3$, the OLNE solution yields a single eutrophic steady
state, associated with significantly lower welfare relative to the
COOP and FBNE solutions.

The primary insight from this comparison is that the COOP and FBNE
solutions lead to outcomes that are remarkably similar in terms of
welfare across states and that their steady states are close. This
result was obtained for the one-dimensional case before but has now
been extended to the two-dimensional case. This is particularly important
for the lake model, which was reduced from a two-dimensional to a
one-dimensional model in the previous literature to make it tractable.
In contrast to the FBNE solutions, the OLNE solution exhibits markedly
different dynamics, with the lake\textquoteright s long-run state\textemdash either
oligotrophic or eutrophic\textemdash being dependent on the initial
conditions in most of the cases examined. Moreover, the OLNE solution
consistently results in lower welfare compared to the COOP benchmark
and the FBNE solution.

\section{Concluding Remarks}

This paper introduces a novel computational method\textemdash the
Strategy Function-Value Function iteration\textemdash for deriving
FBNE solutions in nonlinear differential games. The method was applied
to the canonical lake problem and, to the best of our knowledge, successfully
obtained the FBNE for the first time in the case of two-dimensional
system dynamics. The same approach was also used to compute the cooperative
solution. In addition, we developed a new numerical technique for
solving boundary value problems in order to compute the OLNE solution.
As summarized in Table 1, this framework enables the computation of
the full spectrum of equilibrium outcomes for both one-dimensional
and two-dimensional versions of the lake problem, with two or three
strategic agents.

Our results indicate that the COOP and FBNE solutions are remarkably
close in terms of both steady states and welfare. This finding carries
important policy implications: if agents are incentivized or compelled
to follow the FBNE strategy, outcomes that approximate the cooperative
optimum welfare may be achieved without requiring regulation. We emphasize
that the \textquotedblleft closeness\textquotedblright{} between FBNE
and COOP outcomes is based on numerical approximations and does not
imply exact equivalence. By contrast, the OLNE solution generally
results in significantly lower welfare, highlighting the potential
cost of open-loop behavior.

In open-loop cases, regulatory interventions could be implemented
to steer the system closer to the cooperative outcome. These could
take the form of: (i) price-based instruments, derived from the difference
between the gradient of the value function under the COOP solution
and the costate variables associated with the OLNE; or (ii) quantity-based
instruments, using phosphorus loading trajectories consistent with
the COOP outcome. Our numerical solutions provide these policy-relevant
trajectories across the full state space.

The lake problem served as a testbed for demonstrating the applicability
and effectiveness of the SFVF method. However, the method could be
broadly applicable to other resource management problems that can
be formulated as nonlinear differential games. One promising direction
for future research is the management of fisheries characterized by
predator-prey dynamics\textemdash a two-dimensional system with nonlinear
interactions and separate control variables for harvesting predator
and prey populations. Such a framework would allow further exploration
of cooperative, OLNE, and FBNE solutions under more complex ecological-economic
interactions.

\pagebreak{}

\newpage{}
\begin{doublespace}

\section*{Appendix A: Figures for Three Agents and Accuracy Check}
\end{doublespace}

\begin{doublespace}
\global\long\def\thefigure{A.\arabic{figure}}%
 \setcounter{figure}{0} 
\global\long\def\thesection{A.\arabic{section}}%
 \setcounter{section}{0} 
\global\long\def\thetable{A.\arabic{table}}%
 \setcounter{table}{0} 
\global\long\def\theequation{A.\arabic{equation}}%
 \setcounter{equation}{0}
\global\long\def\thepage{A.\arabic{page}}%
\setcounter{page}{1}
\end{doublespace}

\begin{figure}[H]
\begin{centering}
\includegraphics[width=1\textwidth]{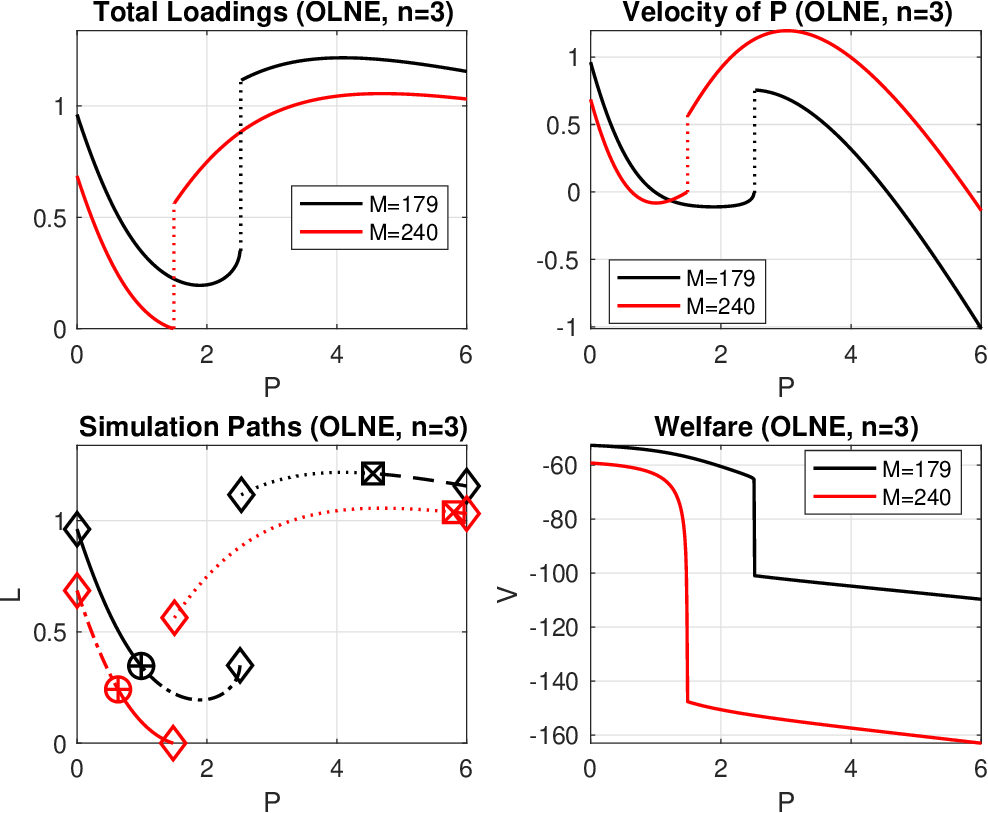}
\par\end{centering}
\caption{\label{fig:OLNE-n3-1D}Solution of One-dimensional Open-loop Nash
Equilibrium when $n=3$}
\end{figure}

\begin{figure}
\begin{centering}
\includegraphics[width=1\textwidth]{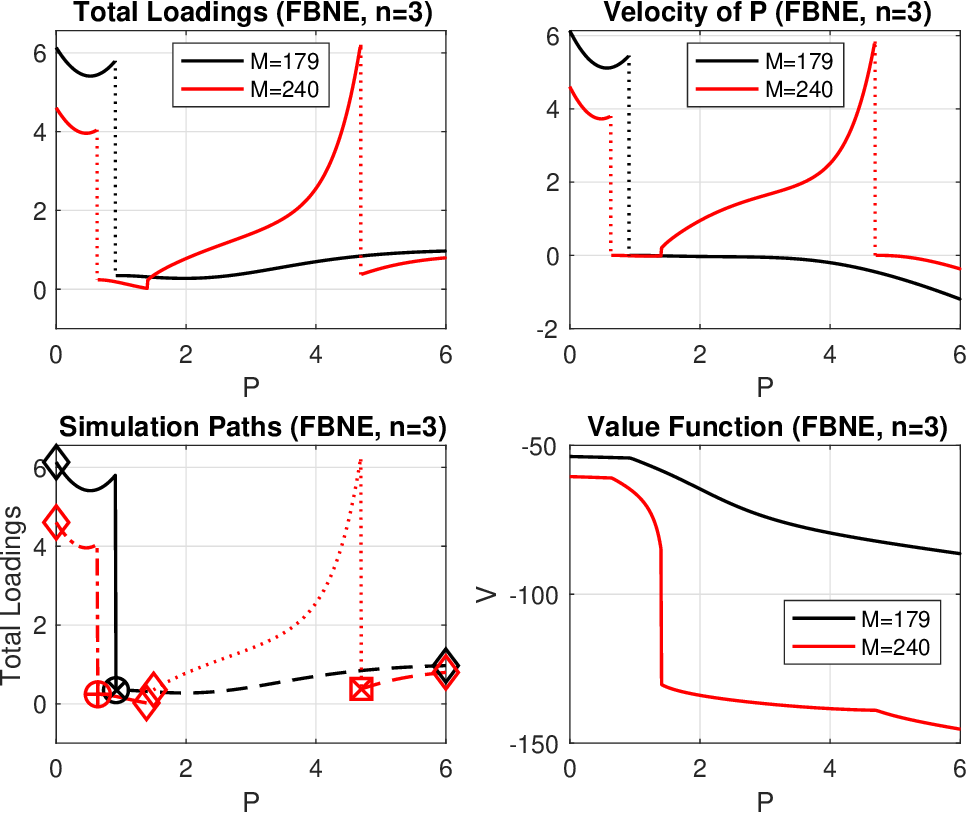}
\par\end{centering}
\caption{\label{fig:FBNE-n3-1D}Solution of One-dimensional Feedback Nash Equilibrium
when $n=3$.}
\end{figure}

\begin{figure}
\begin{centering}
\includegraphics[width=1\textwidth]{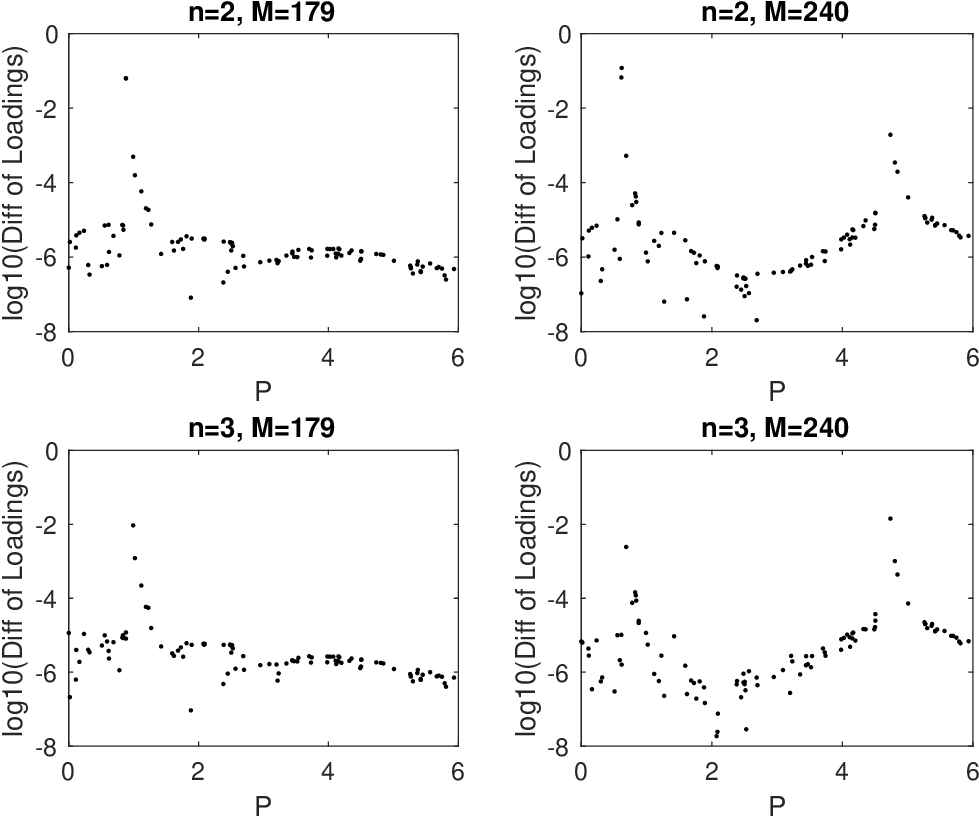}
\par\end{centering}
\caption{\label{fig:Accuracy-FBNE1D}Accuracy of the Solution of One-dimensional
Feedback Nash Equilibrium}
\end{figure}

\begin{figure}
\begin{centering}
\includegraphics[width=0.5\textwidth]{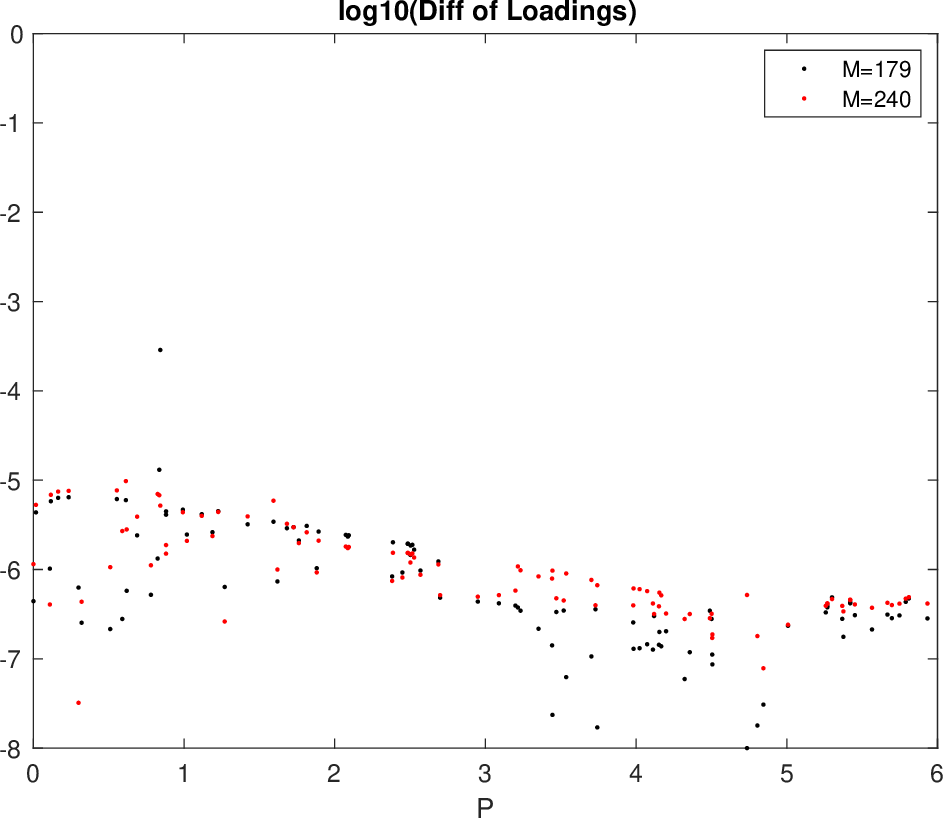}
\par\end{centering}
\caption{\label{fig:Accuracy-COOP1D}Accuracy of the Cooperative Solution of
One-dimensional Lake Problem}
\end{figure}

\begin{figure}
\begin{centering}
\includegraphics[width=1\textwidth]{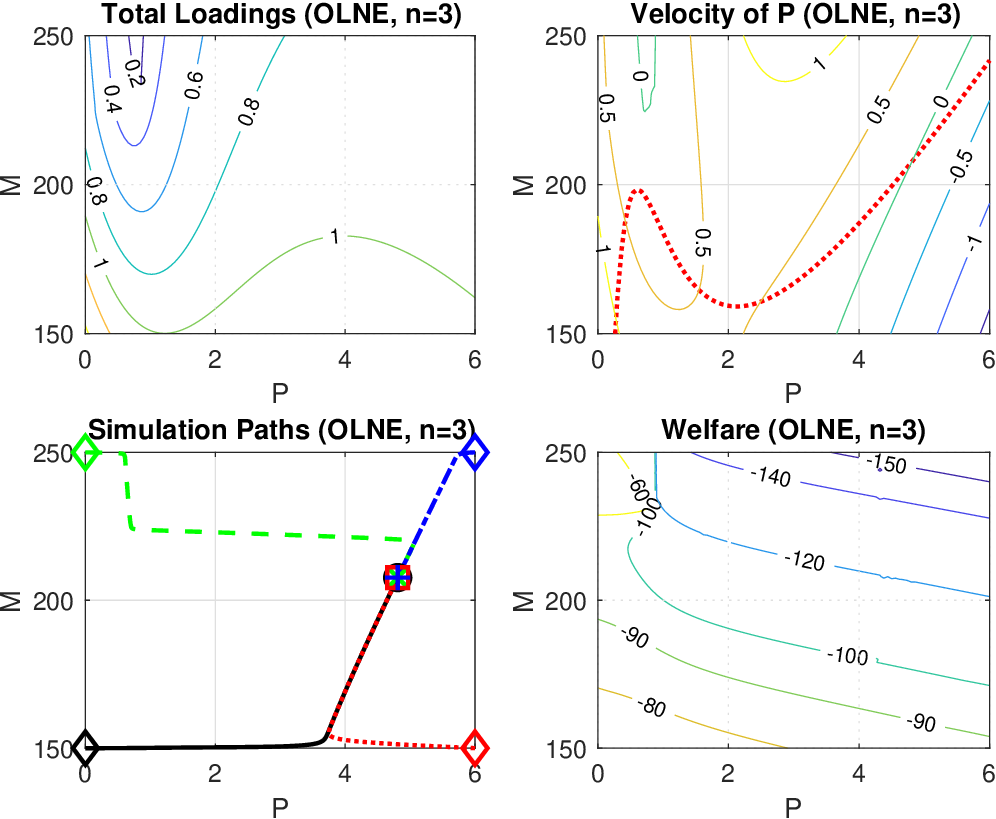}
\par\end{centering}
\caption{\label{fig:OLNE-n3}Solution of Two-dimensional OLNE when $n=3$.
The red dotted line on the top-right panel represents the isocline
$\dot{M}=0$.}
\end{figure}

\begin{figure}
\begin{centering}
\includegraphics[width=1\textwidth]{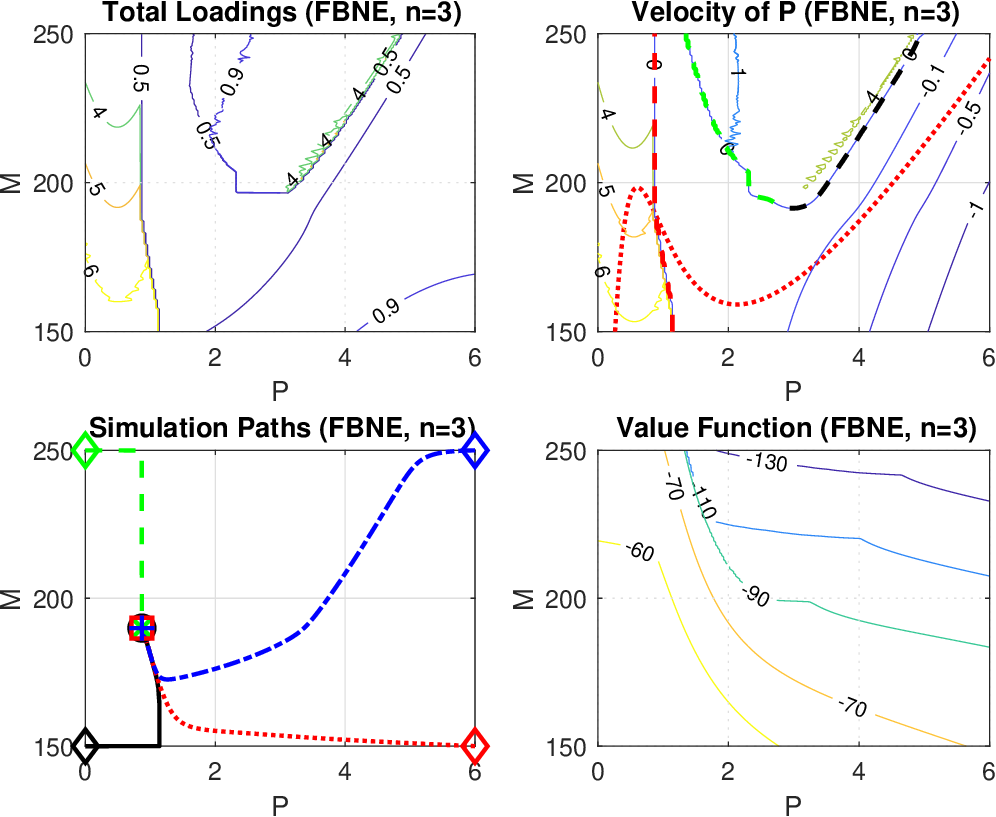}
\par\end{centering}
\caption{\label{fig:FBNE-n3}Solution of Two-dimensional Feedback Nash Equilibrium
when $n=3$. The red dotted line on the top-right panel represents
the isocline $\dot{M}=0$.}
\end{figure}

\begin{figure}
\begin{centering}
\begin{tabular}{cc}
\includegraphics[width=0.5\textwidth]{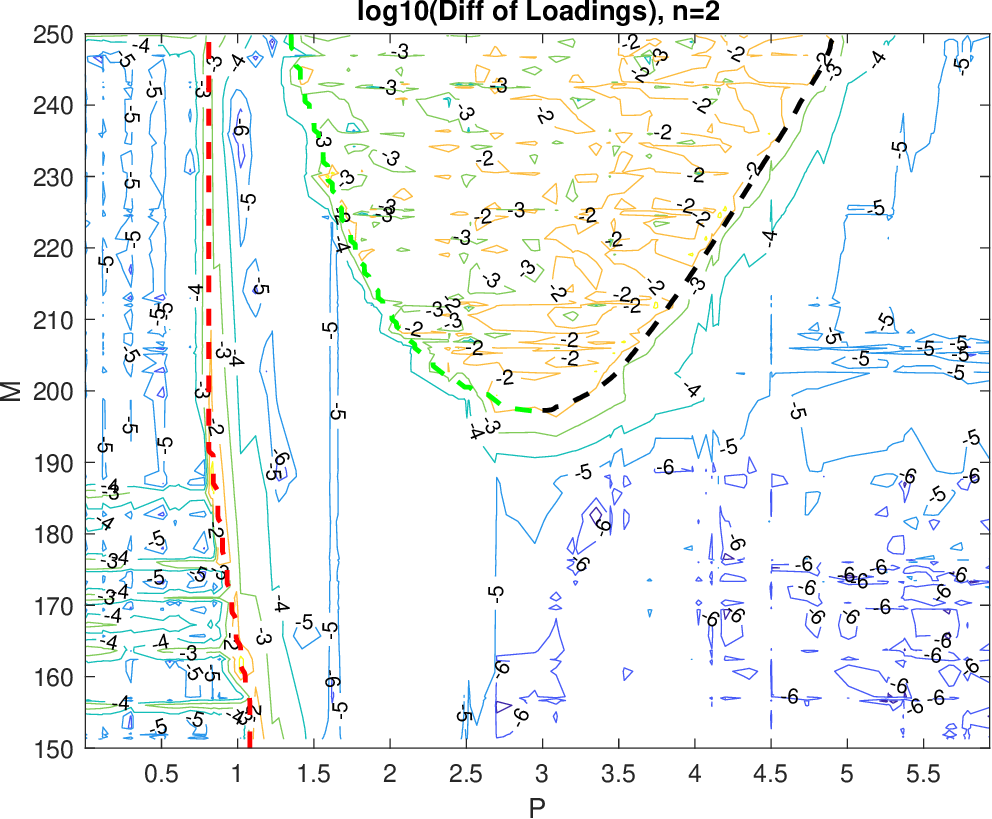} & \includegraphics[width=0.5\textwidth]{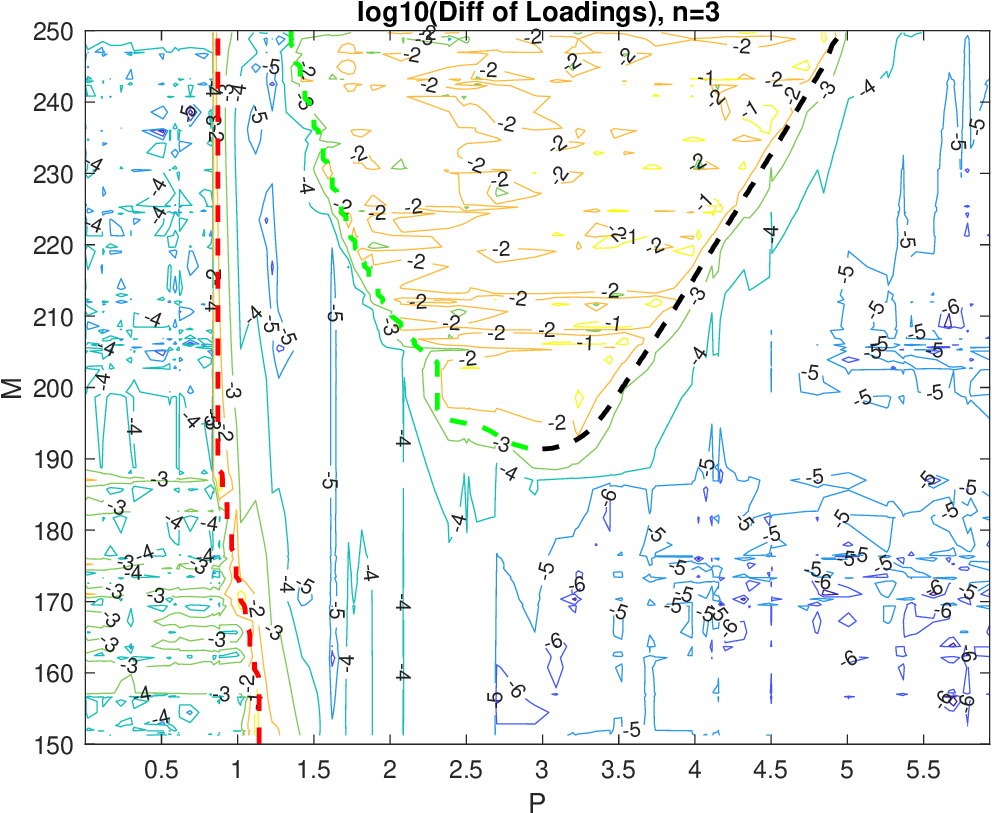}\tabularnewline
\end{tabular}
\par\end{centering}
\caption{\label{fig:Accuracy-FBNE2D}Accuracy of the Solution of Two-dimensional
Feedback Nash Equilibrium}
\end{figure}

\begin{figure}
\begin{centering}
\includegraphics[width=0.5\textwidth]{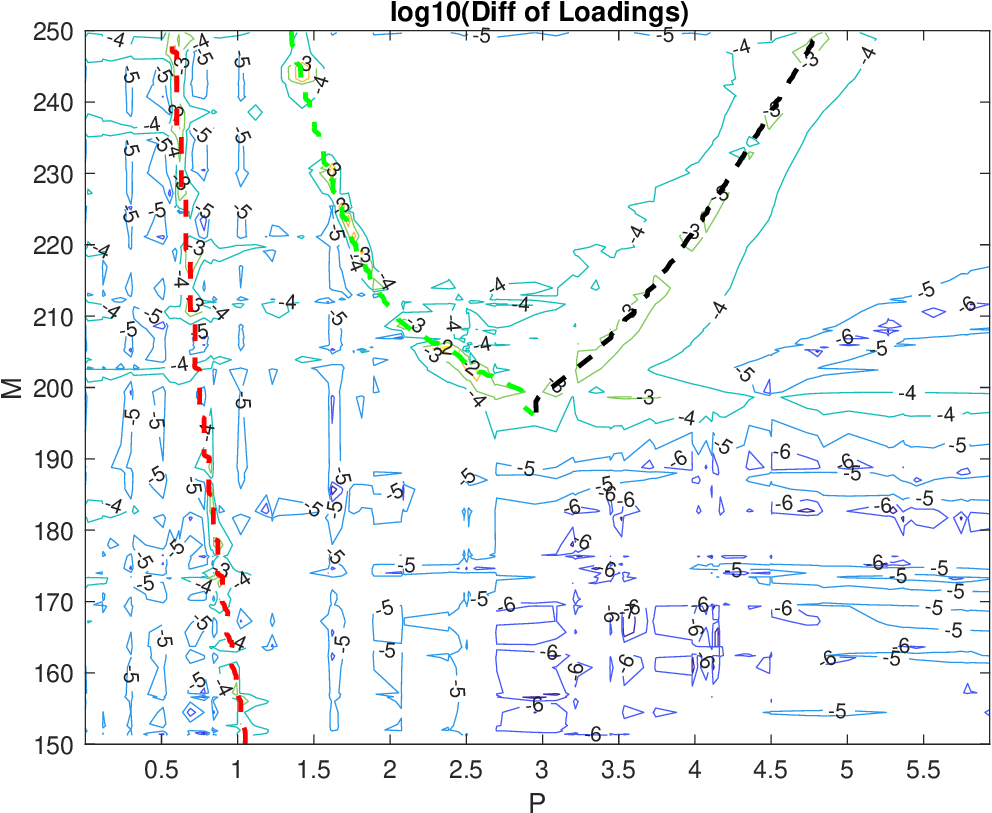}
\par\end{centering}
\caption{\label{fig:Accuracy-COOP2D}Accuracy of the Cooperative Solution of
Two-dimensional Lake Problem}
\end{figure}
\newpage{}

\begin{doublespace}
\global\long\def\thefigure{B.\arabic{figure}}%
 \setcounter{figure}{0}
\end{doublespace}

\section*{Appendix B: Additional Example}

\begin{figure}
\begin{centering}
\includegraphics[width=1\textwidth]{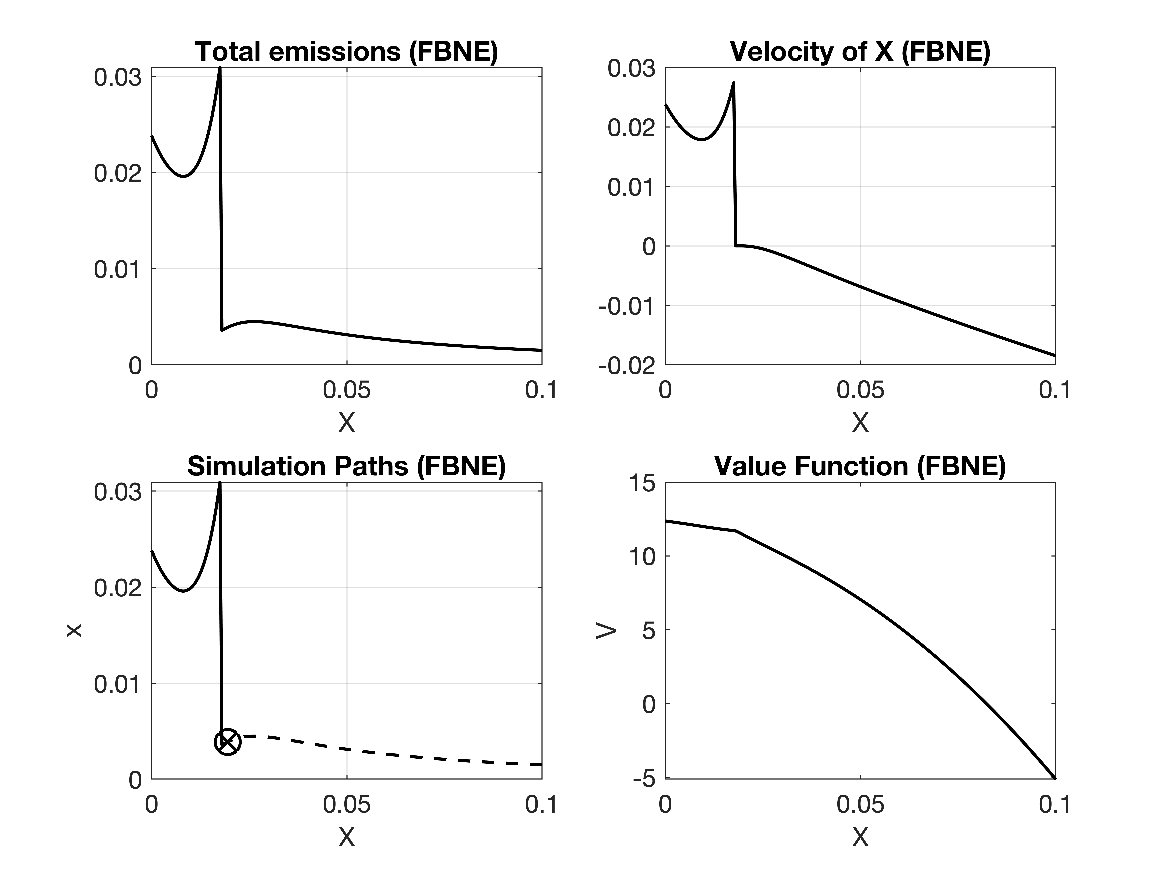}
\par\end{centering}
\caption{\label{fig:Wirl_sol}Numerical Solution of the Feedback Nash Equilibrium
of Wirl's (2007) Model with $a<1/n$}
\end{figure}

\begin{figure}
\begin{centering}
\includegraphics[width=0.5\textwidth]{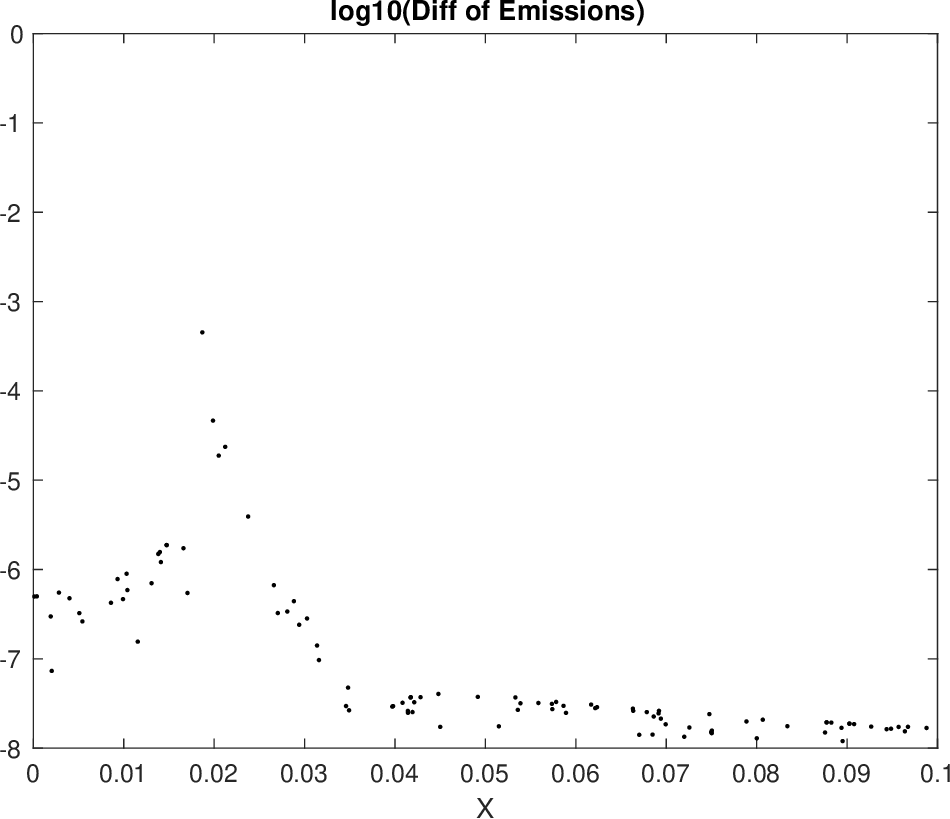}
\par\end{centering}
\caption{\label{fig:Accuracy-of-Wirl}Accuracy of Numerical Solution of the
Feedback Nash Equilibrium of Wirl's (2007) Model with $a<1/n$}
\end{figure}

\begin{figure}
\begin{centering}
\includegraphics[width=0.5\textwidth]{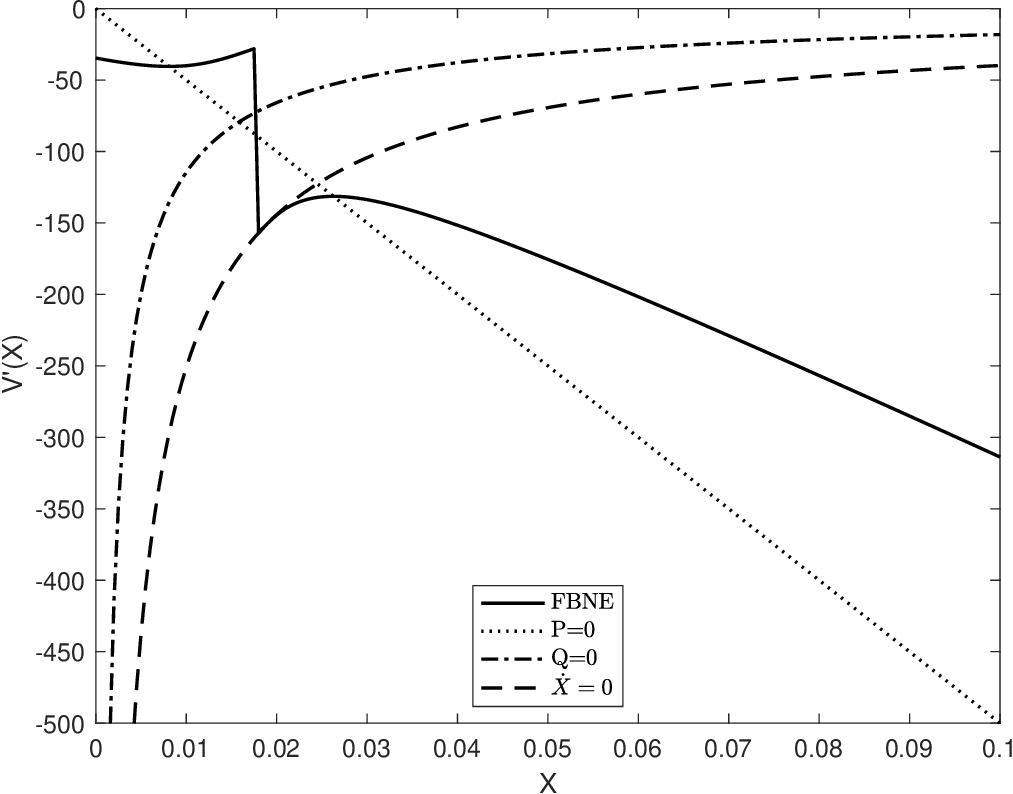}
\par\end{centering}
\caption{\label{fig:Derivative-of-Wirl}Derivative of the Value Function of
Wirl's (2007) Model with $a<1/n$}
\end{figure}

Wirl (2007) discussed multiplicity of symmetric FBNE by analyzing
the following model: 
\begin{equation}
V(X_{0})=\max_{x_{i}(\cdot)}\int^{\infty}_{0}e^{-rt}\left[\frac{1}{a}(x_{i}(t)-\underline{x})^{a}-\frac{c}{2}X(t)^{2}\right]dt\label{eq:FBNE-model-1D-1}
\end{equation}
subject to the transition law of stock $X$ with emissions $x_{i}$
and strategy function $\phi_{j}$:
\[
\dot{X}=x_{i}+\sum_{j\neq i}\phi_{j}(X(t))-\delta X(t)
\]
for players $i=1,...,n$. Under the assumption that the strategy function
$\phi_{j}$ is continuous, Wirl (2007) proved that there is a unique
and singular equilibrium when $a<1/n$. Here we relax the assumption
and apply our SFVF algorithm to solve the model. The parameter values
are $n=2$, $a=0.2$, $r=0.1$, $\underline{x}=0$, $c=1500$, and
$\delta=0.2$. Figure \ref{fig:Wirl_sol} shows that the computed
strategy function is discontinuous and the solution has similar pattern
of our one-dimensional lake model's FBNE with $M=179$. Figure \ref{fig:Accuracy-of-Wirl}
shows that the FBNE solution is highly accurate at the most of randomly
chosen 100 states, except for several states near the jump location
of the strategy function. At the location, $V'$ does not exist, which
leads to relatively larger differences in its vicinity. Figure \ref{fig:Derivative-of-Wirl}
displays the derivative of the value function to compare with Figure
3 of Wirl (2007), where $P$ and $Q$ are defined in Wirl (2007),
that is, $P=0$ and $Q=0$ represent the curves $V'=-cX/(r+\delta)$
and $V'=-\left[(1-a)(n\underline{x}-\delta X)/(an-1)\right]^{a-1}$,
respectively. The figure shows that when $X>n\underline{x}/\delta=0$,
our $V'$ has the same qualitative pattern of Figure 3 of Wirl (2007)
except at the neighborhood of the jump location, because we allow
a discontinuous strategy function and the jump makes the equilibrium
cross $\{Q=0\}$ and reach $\{\dot{X}=0\}$. 

\end{document}